\documentclass[fleqn,usenatbib]{mnras}

\usepackage{newtxtext,newtxmath}
\usepackage[T1]{fontenc}
\usepackage{ae,aecompl}

\DeclareRobustCommand{\VAN}[3]{#2}
\let\VANthebibliography\thebibliography
\def\thebibliography{\DeclareRobustCommand{\VAN}[3]{##3}\VANthebibliography}

\usepackage{graphicx}	% Including figure files
\usepackage{amsmath}	% Advanced maths commands
\usepackage{array}
\usepackage[version=4]{mhchem}
\usepackage{lineno}
\usepackage{amsfonts}
\usepackage{physics}

\usepackage{siunitx}
\usepackage{booktabs}
\newcolumntype{A}{>{\centering\arraybackslash} m{0.13\textwidth} }
\newcolumntype{B}{>{\centering\arraybackslash} m{0.10\textwidth} }
\newcolumntype{C}{>{\centering\arraybackslash} m{0.08\textwidth} }

\makeatletter
\re@DeclareMathSymbol{\varepsilon}{\mathord}{lettersA}{34}
\re@DeclareMathSymbol{\delta}{\mathord}{lettersA}{14}
\makeatother

\title[Pebble accretion in the inner Solar System]{Effects of pebble accretion on the growth and composition of planetesimals in the inner Solar System}

\author[J. Mah et al.]{
J. Mah,$^{1,2}$\thanks{E-mail: mah@mpia.de}
R. Brasser,$^{3}$
A. Bouvier$^{4}$
and S. J. Mojzsis$^{3,5,6}$
\\
$^{1}$Max-Planck-Institut f\"{u}r Astronomie, K\"{o}nigstuhl 17, D-69117 Heidelberg, Germany\\
$^{2}$Earth Life Science Institute, Tokyo Institute of Technology, Tokyo 152-8550, Japan\\
$^{3}$Origins Research Institute, Research Centre for Astronomy and Earth Sciences, H-1112 Budapest, Hungary\\
$^{4}$Bayerisches Geoinstitut, Universit\"{a}t Bayreuth, D-95440 Bayreuth, Germany\\
$^{5}$Department of Lithospheric Research, University of Vienna, A-1090 Vienna, Austria\\
$^{6}$Department of Geological Sciences, University of Colorado Boulder, Boulder, CO 80309-0399, USA\\
}

\date{Accepted 2021 December 22. Received 2021 December 22; in original form 2021 September 7}

\pubyear{2021}

\begin{document}
\label{firstpage}
\pagerange{\pageref{firstpage}--\pageref{lastpage}}
\maketitle

\begin{abstract}
Recent work has shown that aside from the classical view of collisions by increasingly massive planetesimals, the accretion of mm- to m-sized `pebbles' can also reproduce the mass-orbit distribution of the terrestrial planets. Here, we perform {\it N}-body simulations to study the effects of pebble accretion onto growing planetesimals of different diameters located in the inner Solar System. The simulations are run to occur during the lifetime of the gas disc while also simultaneously taking Jupiter’s growth into account. We find that pebble accretion can increase the mass in the solid disc by at least a few times its initial mass with reasonable assumptions that pebbles fragment to smaller-sized grains at the snow line and that gas-disc-induced orbital migration effects are in force. Such a large contribution in mass by pebbles would seem to imply that the isotopic composition of the inner Solar System should be similar to the pebble source (i.e. outer Solar System). This implication appears to violate the observed nucleosynthetic isotopic dichotomy of the sampled Solar System. Thus, pebble accretion played little or no role in terrestrial planet formation.
\end{abstract}


%%%%%%%%%%%%%%%%%%%%%%%%%%%%%%%%%%%%%%%%%%%%%%%%%%

%%%%%%%%%%%%%%%%% BODY OF PAPER %%%%%%%%%%%%%%%%%%

\section{Introduction}
The field of terrestrial planet formation is gradually approaching a clearer understanding of the appropriate initial conditions to form the terrestrial planets with the aid of progressively high-resolution simulations performed on fast computers. In retrospect, the Classical model \citep{Chambers2001} was built upon the results of early analytical derivations \citep[e.g.][]{Wetherill1980,WetherillStewart1989} and numerical simulations \citep{KokuboIda1995,KokuboIda1996,KokuboIda1998} of the evolution of a self-gravitating solid disc of planetesimals. Several planetesimals in the solid disc will grow much faster than the rest -- so-called runaway growth -- and produce a bimodal population in the disc comprising larger-sized planetary embryos and smaller-sized planetesimals \citep{KokuboIda1998}. These embryos continue growing by accreting nearby planetesimals until they exhaust all the planetesimals in their feeding zones. Eventually it is thought that the embryos collide and merge among themselves to finally form the terrestrial planets. It is commonly known, however, that the Classical model runs into the problem that it cannot account for Mars' relatively small mass. Later proposals to improve on the Classical model include the narrow annulus model \citep{Hansen2009}, the Grand Tack model \citep{Walshetal2011}, the low-mass asteroid belt or depleted disc model \citep{Izidoroetal2014,Izidoroetal2015} and the early outer Solar System instability model \citep{Clementetal2018,Clementetal2019,Clementetal2021}.

The dynamical models mentioned above are based on the conventional understanding that the terrestrial planets evolved to their current size through a series of accretion processes involving increasingly massive planetary bodies. There is now an emerging view, however, which postulates that the growth of the terrestrial planets could in fact be a consequence of the accretion of millimetre- to metre-sized bodies dubbed `pebbles', in direct conflict to the Classical view. Although the {\it pebble accretion} model \citep{OrmelKlahr2010,LambrechtsJohansen2012} was originally developed to resolve a problem related to the formation of the core of the gas giants in the outer Solar System \citep[e.g.][]{Levisonetal2015giantpl}, several recent studies \citep[e.g.][]{Levisonetal2015,Lambrechtsetal2019,Johansenetal2021} have applied this model to the terrestrial planet region and found positive results that encourage further investigations on the plausibility of this formation pathway for the terrestrial planets.

Pebbles, observed to be abundant in protoplanetary discs \citep[e.g.][]{Testietal2003,Wilneretal2005,Rodmannetal2006}, are thought to have formed via the coagulation of smaller-sized dust particles in the outer regions of the disc \citep{LambrechtsJohansen2014,Idaetal2016}. They tend to spiral towards the central star under the influence of gas drag because dynamical friction with the surrounding gas causes them to lose angular momentum \citep{Weidenschilling1977}. During their passage through the protoplanetary disc, these pebbles should encounter planetesimals in the inner Solar System and some of the pebbles will accrete onto the planetesimals.

The work of \citet{Levisonetal2015} investigated the growth and evolution of planetesimals distributed within 2.7~au of the Sun to a flux of inward-drifting pebbles that are continuously generated in the same region over a period of 2~Myr. Since in their simulations the pebbles are formed continuously in the inner Solar System, the surface density of the gas was chosen to be about 5 times the minimum mass solar nebula \citep[MMSN;][]{Hayashi1981}, a value deemed sufficient to sustain continuous pebble formation in the disc. The authors found that the planetesimals closer to the Sun accrete more pebbles and grow to larger sizes compared to planetesimals further away because the pebble accretion cross-section has a semi-major axis dependence resulting from their assumptions on the Stokes number -- a number that quantifies how well the pebbles are coupled to the gas. Mars' smaller size compared to Earth and Venus can therefore be easily reproduced because it formed from the mergers of smaller-sized planetesimals. It is perhaps more accurate to describe the contribution of the pebble accretion mechanism in this model as helping to generate the appropriate initial mass-semi-major axis distribution of planetesimals and planetary embryos that would naturally reproduce the masses of the terrestrial planets via subsequent collisions and mergers. The key assumption of this work is that the pebbles contribute to the mass of planetesimals formed locally in the inner Solar System. This assumption was required because the pebble flux necessary to form the cores of the giant planets \citep{Levisonetal2015giantpl} is too great to form the terrestrial planets. The assumption of local pebble formation over time is, however, peculiar and deviates from the more common assumption used in many studies (including ours) that the pebbles originate in the outer Solar System.

The work of \citet{Lambrechtsetal2019} also looked at the effects of pebble accretion in the inner disc but assumed that the pebbles originate from a region further away. Their simulations start with lunar mass planetary embryos distributed between 0.5 and 3.0~au following an $r^{-2}$ surface density profile. While the gas disc model, pebble accretion formulation and orbital migration prescriptions employed in their work are broadly similar to ours, there are notable differences: the disc temperature and pebble Stokes number $(\mathcal{S}=3\times10^{-3})$ are kept constant. They focused on how the amount of pebble flux flowing through the disc influences the mass-distance distribution of the embryos after gas dissipation and the final planetary system architecture. \citet{Lambrechtsetal2019} demonstrated that there exists a strong dependence of planetary system orbital architecture on the pebble flux. A pebble flux of $\sim 110$ Earth mass $(M_{\rm E})$ per Myr would result in Mars-sized embryos after the gas disc phase and terrestrial planet-like systems after about 100~Myr while a mere doubling of the pebble flux would produce close-in super-Earth-like systems. The authors then suggest that the growth of Jupiter could have restricted the pebble flux to the inner disc and thus put a limit on the final mass of the terrestrial planets.

In recent work, \citet{Johansenetal2021} propose that the planetesimals that went on to become the terrestrial planets (Venus, Earth and Mars) all formed initially near the water snow line, and that these planetesimals subsequently migrated inwards to the current semi-major axis of the terrestrial planets while accreting pebbles along the way. The snow line in their disc model starts between 1.2~au to 2.0~au and migrates inwards with time. The planet formation scenario presented in their work is based on good analytical fits of the growth track equations of \citet{Johansenetal2019}. The masses of Venus and Mars can be reproduced using reasonable values of the disc temperature profile \citep[$T \propto r^{-3/7}$;][]{ChiangGoldreich1997}, Stokes number $(10^{-3}<\mathcal{S}<0.1)$ and pebble flux. The study assumes that these planets' current masses are their masses at the time of gas disc dissipation, and that both Venus and Mars grew entirely from pebbles without giant impacts. For Earth, the maximum mass achieved after the end of the pebble accretion phase (when the gas disc has dissipated) is $0.6~M_{\rm E}$. A purported giant impact with a body of mass $0.4~M_{\rm E}$ is suggested to occur later so that the mass of the Earth can reach its current value. In their model, the terrestrial planets (Venus, Mars and proto-Earth) accreted pebbles of two different compositions in the sequence of (1) non-carbonaceous (specifically, ureilite-like) in the first few Myr, followed by (2) carbonaceous (specifically, CI-like) until the dissipation of the gas disc. Based on earlier studies of calcium and iron isotopes, \citet{Johansenetal2021} assume that $\sim 40\%$ of Earth's composition comes from pebbles that originated in the outer Solar System with an assumed CI-like isotopic composition \citep{Schilleretal2018,Schilleretal2020}. The low mass of Mars can be achieved by assuming that it formed later than the other terrestrial planets or that it stopped accreting earlier \citep{Schilleretal2018}. While the suggestion of planetesimal formation triggered by snow line passage is not new \citep{DrazkowskaAlibert2017}, the only manner in which this can result in planetesimals with a different composition is if the material that condensed at the snow line to become planetesimals has a different composition at a different distance to the Sun, which \citet{Johansenetal2021} argue occurred because of the initial viscous expansion of the disc. Yet, we emphasize here that planetesimal formation at a static snow line is difficult to reconcile with the different isotopic and elemental compositions of the sampled solar system, including Earth and Mars.

The results of prior studies examining the formation of the terrestrial planets in the framework of the pebble accretion model are indeed interesting and they set the stage for further research in this direction. The work of \citet{Levisonetal2015} and \citet{Lambrechtsetal2019} showed that pebble accretion (with a specific value of the pebble mass flux and Stokes number) played a supportive role in the early growth stages of planetesimals in the inner Solar System. In contrast \citet{Johansenetal2021} reported that pebble accretion could potentially be the primary mechanism for the formation of the terrestrial planets in a single stage. This latter hypothesis bypasses the need for an epoch of hierarchical accretion by ever-larger planetesimals and planetary embryos.

Here, we study the effects of pebble accretion in the inner Solar System over the lifetime of the gas disc while simultaneously taking into account the growth of Jupiter. To do this, we run a suite of {\it N}-body simulations with different initial conditions (disc temperature, solid disc mass) with the aim of quantifying the amount of mass delivered to the inner disc by pebble accretion. The pebble accretion formulation we adopt has its foundation on the work of \citet{Idaetal2016} which provides equations for the pebble accretion rate at different locations in a steady state accretion disc while consistently taking into account the various properties of the gas disc (e.g. temperature, surface density, scale height). We chose to use this particular pebble accretion formulation for its simplicity and elegance, and because it does not invoke any additional assumptions \citep[cf.][]{Levisonetal2015,Johansenetal2021}.

The major differences between our work and the studies summarised previously lie in the gas disc model we adopted and the computation and underlying assumptions of the value of the Stokes number. In our simulations, we follow \citet{Idaetal2016} and compute the radius and the Stokes number of the pebbles self-consistently depending on their location in the disc and time. In contrast, the Stokes number used in the works of \citet{Levisonetal2015} and \citet{Lambrechtsetal2019} is fixed while \citet{Johansenetal2021} assume that it lies in a particular range and that the pebbles are mm-sized. Our simulations are in a sense more realistic concerning the treatment of pebbles, but the downside is that a longer computational time is required and that our gas disc model may be too simplistic.
%%%%%%%%%%%%%%%%%%%%%%%%%%%%%%%%%%%

%%%%%%%%%%%%%%%%%%%%%%%%%%%%%%%%%%%
\section{Pebble accretion onto planetesimals}

\subsection{Basic formulation}
\label{sec:pebacc_formula}
Based on the formulation of \citet{OrmelKlahr2010} and \citet{OrmelKobayashi2012} and modified by \citet{Idaetal2016}, the accretion rate of mm to m-sized pebbles onto a planetesimal of mass $M$ and located at distance $r$ from the central star is derived to be
\begin{equation}
    \dot{M} = \min \left[1,\frac{C b^2\sqrt{1+4\mathcal{S}^2}}{4\sqrt{2\pi}\mathcal{S}h_{\rm peb}}\left(1 + \frac{3b}{2\chi\eta}\right)\right]{\dot M_{\rm F}}, 
    \label{eq:dotmpl}
\end{equation}
where $C$ is a constant related to the pebble accretion mode (equation~\ref{eq:const2d3d}), $b$ is the normalised (or reduced) impact parameter (equation~\ref{eq:impactparam}), $\mathcal{S}$ is the Stokes number which determines the degree of coupling between the pebbles and the gas (equation~\ref{eq:taus}), $h_{\rm peb}$ is the normalised pebble scale height (equation~\ref{eq:hpeb}), $\chi$ is a function of the Stokes number (equation~\ref{eq:chi}), $\eta$ is the difference between the gas and Keplerian velocities (equation~\ref{eq:eta}), and $\dot{M}_{\rm F}$ is the pebble mass flux (equation~\ref{eq:mdotf}). In the following we describe each of these quantities in more detail.

The flux of pebbles through the disc at a specific time and location from the star $\dot{M}_{\rm F}$ is one of the key quantities that control the pebble accretion rate. Its temporal evolution depends on the gas accretion rate onto the central star and the formation efficiency of pebbles. For our fiducial disc model it is given by \citep{Idaetal2016}
\begin{equation}
    \dot{M}_{\rm F} = 10^{-3}\alpha_3^{-1}\dot{M}_{*8}\left(\frac{L_*}{L_\odot}\right)^{2/7}\left(\frac{M_*}{M_\odot}\right)\left(\frac{t+t_{\rm off}}{1~{\rm Myr}}\right)^{-1/3}~M_{\rm E}~{\rm yr}^{-1},
    \label{eq:mdotf}
\end{equation}
where $\alpha_3 \equiv \alpha/10^{-3}$, $\alpha$ is the disc viscosity parameter \citep{ShakuraSunyaev1973}, $L_* (L_{\odot})$ is the luminosity of the star (Sun), $M_* (M_{\odot})$ is the mass of the star (Sun) and $t_{\rm off}$ is the timescale at which the planetesimals are assumed to instantaneously form. The other key quantity that influences the pebble accretion rate is the normalised collision impact parameter $b$, which takes the form
\begin{equation}
    b = 2\kappa \frac{R_{\rm H}}{r} \mathcal{S}^{1/3}~\min \left(\sqrt{\frac{3R_{\rm H}}{\chi\eta r}}\mathcal{S}^{1/6},1\right),
    \label{eq:impactparam}
\end{equation}
where $\kappa$ is the reduction factor accounting for cases where pebbles are weakly coupled to the gas resulting in inefficient accretion, $R_{\rm H} = r(M/3M_*)^{1/3}$ is the Hill radius of the planetesimal. The terms in the parentheses encompass the two possible pebble accretion regimes, namely the Bondi regime (left-hand term) where the relative velocity between the planetesimal and a pebble is dominated by the pebble's drift velocity, and the Hill regime (right-hand term) where the relative velocity is dominated by Keplerian shear. The quantity $\chi$ is defined as
\begin{equation}
    \chi = \frac{\sqrt{1+4\mathcal{S}^2}}{1+\mathcal{S}^2}.
    \label{eq:chi}
\end{equation}
The Stokes number itself is given by
\begin{equation}
    \mathcal{S} = \frac{\rho_{\rm peb}R_{\rm peb}}{\rho_{\rm g} H_{\rm g}}\max\left(1,\frac{4R_{\rm peb}}{9\lambda}\right), 
    \label{eq:taus}
\end{equation}
where $\rho_{\rm peb}$ and $R_{\rm peb}$ are the bulk density and the physical radius of a pebble, $\rho_{\rm g}$ is the gas midplane density, the disc scale height $H_{\rm g} = c_{\rm s}/\Omega_{\rm K} \propto T^{1/2}$ is a function of the disc temperature $T$, the sound speed $c_{\rm s}$ and the Keplerian orbital frequency $\Omega_{\rm K} = \sqrt{GM_*/r^3}$, and $\lambda$ is the mean free path of the pebble. The gas midplane density $\rho_{\rm g}$ is related to the disc surface density $\Sigma_{\rm g}$ and the disc scale height $H_{\rm g}$ via
\begin{equation}
    \rho_{\rm g} = \frac{\Sigma_{\rm g}}{\sqrt{2\pi}H_{\rm g}}.
\end{equation}
The pebble mean free path is given by
\begin{equation}
    \lambda = \frac{\mu m_{\rm H}}{\sigma_{\rm H_2} \rho_{\rm g}},
\end{equation}
where $\mu$ is the mean molecular weight of the gas (mostly hydrogen), $m_{\rm H}$ is the mass of a hydrogen atom (mostly that of two protons), and $\sigma_{\rm H_2}$ is the collisional cross section of a hydrogen molecule. The reduction factor $\kappa$ is a function of the Stokes number, expressed as \citep{OrmelKobayashi2012}
\begin{equation}
    \ln \kappa = -\left(\frac{\mathcal{S}}{\mathcal{S}^*}\right)^{0.65},
    \label{eq:kappa}
\end{equation}
where the quantity $\mathcal{S}^*$ is defined as
\begin{equation}
    \mathcal{S}^* = \min \left[2,4 \eta^{-3} \frac{M}{M_*}\right].
    \label{eq:S*}
\end{equation}
The quantity $\eta$ is given by
\begin{equation}
    \eta = \frac{1}{2}h_{\rm g}^2 \left| \dv{\ln P}{\ln r} \right|,
    \label{eq:eta}
\end{equation}
where $h_{\rm g} = H_{\rm g}/r$ is the reduced gas scale height. The remaining quantities in equation~\ref{eq:dotmpl} that we have yet to describe are the constant $C$ that determines the mode of accretion (whether 2D or 3D) and the reduced pebble scale height $h_{\rm peb}$. These quantities are expressed as \citep{Idaetal2016}
\begin{equation}
    C = \min\left(\sqrt{\frac{8}{\pi}}\frac{h_{\rm peb}}{b}, 1\right),
    \label{eq:const2d3d}
\end{equation}
and
\begin{equation}
    h_{\rm peb} = \left(1+\frac{\mathcal{S}}{\alpha}\right)^{-1/2}h_{\rm g}.
    \label{eq:hpeb}
\end{equation}
Pebble accretion begins in 3D mode ($C \sim 1$) when the planetesimal is of low mass and then transitions into the 2D mode when the planetesimal grows larger and its Hill radius becomes larger than the scale height of the pebbles $(R_{\rm H} > h_{\rm peb})$.
%%%%%%%%%%%%%%%%%%%%%%%%%%%%%%%%%%%

%%%%%%%%%%%%%%%%%%%%%%%%%%%%%%%%%%%
\subsection{Application to small planetesimals}
According to \citet{OrmelKlahr2010} pebble accretion causes rapid growth onto planetesimals with diameters of $\sim 200~{\rm km}$ or greater. To understand how pebble accretion works on small(er) bodies the question becomes which of the paths in the $\min()$ statements in equations~\ref{eq:dotmpl},~\ref{eq:impactparam},~\ref{eq:S*} and~\ref{eq:const2d3d} above dominate.

We mentioned previously that the two key quantities controlling the pebble accretion rate are the global pebble flux $\dot{M}_{\rm F}$ and the normalised impact parameter $b$. In the following, we will demonstrate that the reduction factor $\kappa$ (equation~\ref{eq:kappa}) turns out to be important as well. The quantity $\mathcal{S}^*$ (equation~\ref{eq:S*}) sets the boundary between the so-called {\it settling regime}, in which the cross section for accretion is very large up to the Hill radius \citep{OrmelKobayashi2012}, and the {\it geometric regime}, in which the cross section is just the surface area of the planetesimal \citep{OrmelKlahr2010}. The two regions are smoothly connected with the parameter $\kappa$, whose functional form is a fit that \citet{OrmelKlahr2010} obtained from numerical experiments. The critical Stokes number that separates the settling regime from the geometric regime results in $\mathcal{S}^* < 2$. At around 1~au, this translates to a critical planetesimal-to-star mass ratio of $M/M_* < \frac{1}{2}\eta^3 \lesssim 1.4\times10^{-9}$ which corresponds to a diameter of $D=1200~{\rm km}$ for a bulk density of $\rho = 3000~{\rm kg~m}^{-3}$ for the planetesimal, assuming nominal disc quantities (see Appendix~\ref{sec:appendix_discmodel} for a description of our disc model). This critical value of $M/M_*$ has a distance dependence. For example, at 3~au $\mathcal{S}^* < 2$ when the planetesimal mass is $M/M_* < 3.5\times10^{-9}$, or equivalent to a diameter of 1700~km. As such, when a planetesimal at 3~au has a diameter smaller than about twice that of Ceres, or that it is roughly 1200~km if it is located at 1~au, the quantity $\mathcal{S}^*$ can become smaller than 2, meaning that for these cases we are no longer in the settling regime and are instead approaching the geometric regime where the accretion cross section reduces to the surface area of the planetesimal. This has important implications with regards to the pebble accretion rate and efficiency of planetesimals below a critical diameter at a given semi-major axis.

In the asteroid belt, the typical value for the pebble Stokes number $\mathcal{S}$ with our model parameters is 0.1 and the quantity $\mathcal{S}^*$ is often equal to 2 \citep{Idaetal2016}. For these values, the reduction factor $\kappa = 0.87$. Suppose now that we want to study accretion onto the parent body of the Howardite-Eucrite-Diogenite (HED) meteorites, which, despite existing controversies, we here assume to be the asteroid 4 Vesta \citep{McCordetal1970,ConsolmagnoDrake1977,BinzelXu1993,Keil2002,McSweenetal2013}. Vesta's diameter is $D \sim 500~{\rm km}$ and $M/M_* = 1.3\times10^{-10}$. For our nominal disc model at 2.5~au (Vesta's modern orbital distance), $\eta \sim 3\times10^{-3}$ and $\mathcal{S}^* \sim 0.02$. If the Stokes number is the typical value of 0.1 then $\kappa = 0.06$, but if the Stokes number is higher then $\kappa$ will be lower, resulting in a much lower rate of accretion than when the impact parameter is almost equal to the Hill radius. Pebble accretion onto the proposed (ordinary chondrite, OC) H-chondrite parent body 6 Hebe \citep{Gaffeyetal1993,GaffeyGilbert1998,Binzeletal2004,Binzeletal2019}, with $D \sim 200~{\rm km}$ and $M/M_* \lesssim 7\times10^{-12}$, is much less efficient yet again because the quantities $\mathcal{S}^*$ and $\kappa$ work out to be $\lesssim 10^{-3}$ and $\lesssim  2\times10^{-9}$, respectively. Thus, because of the $\kappa$ factor, which is related to how the pebbles accrete in the settling regime \citep{OrmelKlahr2010,OrmelKobayashi2012}, accretion onto small objects with $D \sim 200~{\rm km}$ could become {\it extremely inefficient}. However, the impact parameter $b$ must always be equal to or greater than the radius of the planetesimal for it to be physically meaningful. For a planetesimal with bulk density $\rho = 3000~{\rm kg~m}^{-3}$ we have $R/R_{\rm H} \sim 5.7\times10^{-3}$ so that the relative impact parameter in terms of the Hill radius cannot go much below this value. The revised impact parameter therefore should read
\begin{equation}
        b=\max \left[2\kappa \frac{R_{\rm H}}{r} \mathcal{S}^{1/3}~\min \left(\sqrt{\frac{3R_{\rm H}}{\chi\eta r}}\mathcal{S}^{1/6},1\right),\frac{R}{r}\right],       
\end{equation}
where $R$ is the radius of the planetesimal. This shows that the accretion rate of pebbles cannot become arbitrarily low, but that small planetesimals are still severely disadvantaged in terms of their accretion rate over their larger brethren.
%%%%%%%%%%%%%%%%%%%%%%%%%%%%%%%%%%%

%%%%%%%%%%%%%%%%%%%%%%%%%%%%%%%%%%%
\subsection{The role of the snow line and fragmentation}
\label{sec:pebacc_fragmentation}
\citet{Morbidellietal2016} suggest that the pebbles lose their volatiles at the snow line and could fragment into mm-sized grains akin to chondrules. This fragmentation, if confirmed by mechanical arguments, lowers the Stokes number (equation~\ref{eq:taus}). The Stokes number can either be $\mathcal{S} \propto R_{\rm peb}$, called the Epstein regime, or $\mathcal{S} \propto R_{\rm peb}^2$, called the Stokes regime, depending on the distance to the star and the properties of the gas disc. The transition between the two regimes occurs at \citep{Idaetal2016}
\begin{equation}
    r_{\rm ES} = 2.2\rho_{\rm peb}^{-7/26}\left(\frac{L_*}{L_\odot}\right)^{-3/13}\left(\frac{M_*}{M_\odot}\right)^{17/26}\left(\frac{\dot{M}_{\rm F4}^{1/3}\dot{M}_{*8}}{\alpha_3}\right)^{21/52}~{\rm au}, 
\end{equation}
where we additionally define $\dot{M}_{\rm F4} \equiv \frac{\dot{M}_{\rm F}}{10^{-4}~M_{\rm E}~{\rm yr}^{-1}}$ and $\dot{M}_{*8} \equiv \frac{\dot{M}_*}{10^{-8}~M_\odot~{\rm yr}^{-1}}$. For nominal gas disc temperature and surface density at 1~au, the pebble mean free path $\lambda \sim 2~{\rm cm}$. Inside the snow line, where fragmentation could occur, the pebble radius is assumed to be $R_{\rm peb} \sim 1~{\rm mm}$ \citep{Morbidellietal2016} and so in this case $\mathcal{S} \propto R_{\rm peb}$ because $R_{\rm peb} \ll \lambda$ (equation~\ref{eq:taus}). Assuming the pebble density as $\rho_{\rm peb}=1000~{\rm kg~m}^{-3}$, the Stokes number at 1~au for nominal disc parameters is approximately $\mathcal{S}\sim 0.2$, but after fragmentation it is lowered to $\mathcal{S} \sim 10^{-4}$ \citep{BrasserMojzsis2020}.

With the reduction of the Stokes number at 1~au from $\mathcal{S} \sim 0.2$ to $\mathcal{S} \sim 10^{-4}$ as a consequence of potential fragmentation, pebble accretion only becomes inefficient once $\mathcal{S}^* \lesssim 2\times10^{-5}$, i.e., when $M/M_* \lesssim 10^{-14}$, which corresponds to a planetesimal diameter of $D\lesssim 50~{\rm km}$. This critical diameter is lower than that derived previously for the case where fragmentation is not considered (cf. $D \sim 200~{\rm km}$). Thus, with the inclusion of fragmentation effects the range of planetesimal diameters for which pebble accretion can proceed efficiently becomes wider. Importantly, the accretion rate also depends strongly on the Stokes number, and a lower Stokes number actually results in a much slower overall rate of accretion with all other parameters being equal. Nevertheless, due to the debilitating effect of $\kappa$, accretion onto small planetesimals by fragmented pebbles is still faster than by intact pebbles beyond the snow line. As such, if fragmentation occurred, pebble accretion onto planetesimals with a diameter comparable to the H-chondrite parent body $(D \sim 200~{\rm km};~{\rm 6~Hebe})$ or smaller can still proceed, albeit with low efficiency, but only as long as they remain inside the snow line. To summarise, the temporal evolution of the snow line plays a critical role in the growth rate of planetesimals in the inner disc from the inward spiralling pebbles. 
%%%%%%%%%%%%%%%%%%%%%%%%%%%%%%%%%%%

%%%%%%%%%%%%%%%%%%%%%%%%%%%%%%%%%%%
\subsection{Analytical computation of pebble accretion rate}
\label{sec:pebacc_analytical}
\begin{figure*}
    \centering
	\includegraphics[width=\textwidth]{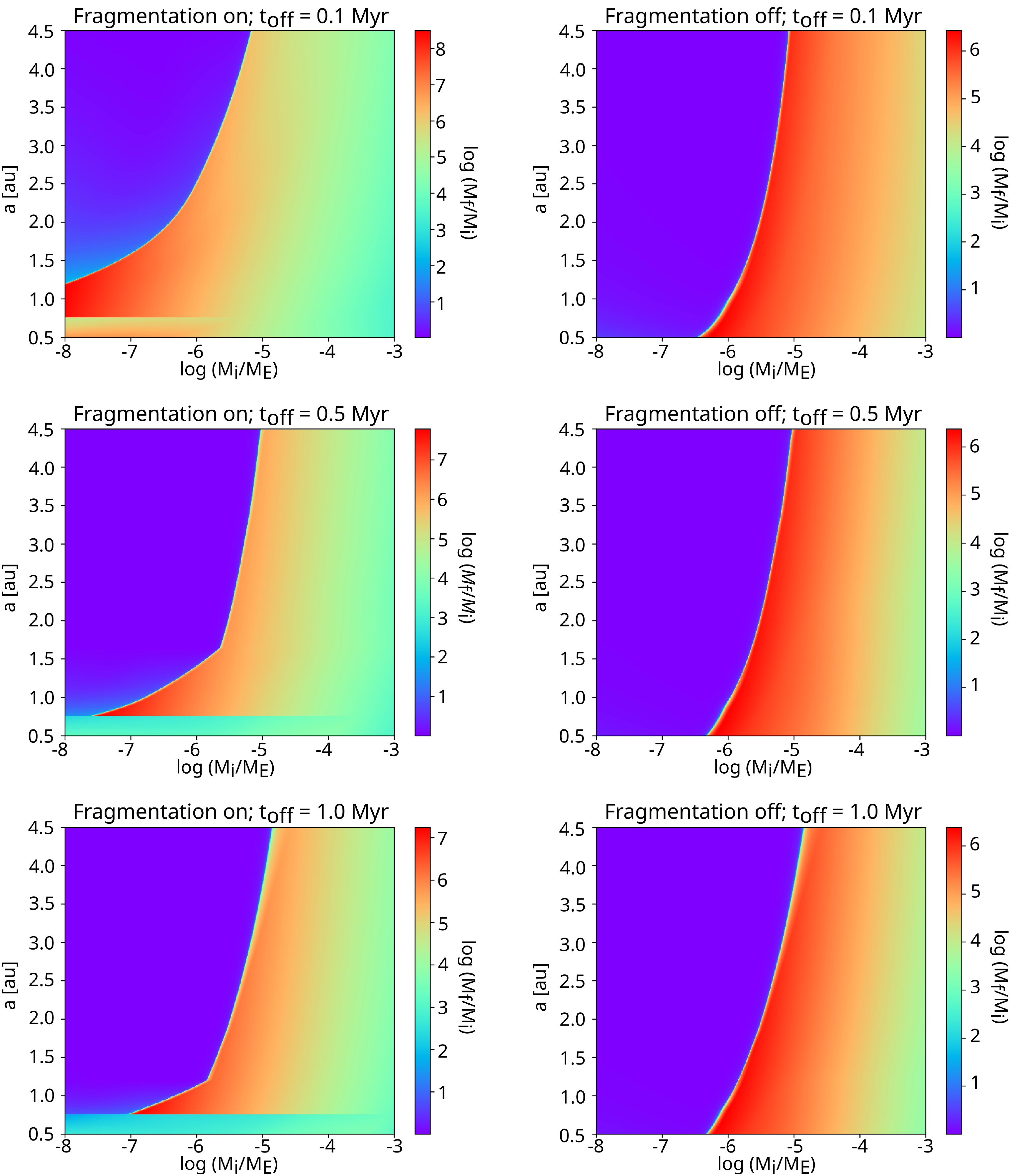}
    \caption{Contour plot depicting the log of the mass ratio of the planetesimal $\log (M_{\rm f}/M_{\rm i})$ as a function of its initial mass $M_{\rm i}$ and semi-major axis $a$, with and without pebble fragmentation effects at the snow line. The top row assumes all planetesimals form very soon after the disc at $t_{\rm off} = 0.1~{\rm Myr}$, the middle row assumes planetesimal formation after 0.5~Myr and the bottom row after 1~Myr. Regardless of our choice of parameters (initial planetesimal size, planetesimal semi-major axis, planetesimal formation timing, inclusion/exclusion of fragmentation effects), pebble accretion onto small planetesimals in the asteroid belt is very inefficient. The planetesimal diameter can be computed from $\log (D/{\rm 1~km})= \frac{1}{6}\log (M/1~M_{\rm E})+3.597$, so that $M/M_{\rm E} =10^{-5}$ corresponds to $D=580~{\rm km}$.}
    \label{fig:pebacc_contour}
\end{figure*}

We combine all of these aspects to show a global accretion map in Fig.~\ref{fig:pebacc_contour}, which is a contour plot of $\log (M_{\rm f}/M_{\rm i})$, i.e., the log of the final mass of a planetesimal divided by its initial mass. We chose to plot this instead of the final mass to better show the total mass growth. The map was created by integrating equation~\ref{eq:dotmpl} for a single planetesimal located at a fixed semi-major axis as a function of time either for at most 4~Myr \citep[approximate lifetime of the gas disc;][]{Wangetal2017} or until the pebble isolation mass \citep{Lambrechtsetal2014} was reached. The pebble isolation mass of a planetary body is the mass above which pebble accretion stops due to the planetary body causing perturbations in the disc that in turn halt the flow of pebbles \citep{Lambrechtsetal2014}. Each panel of Fig.~\ref{fig:pebacc_contour} is made up of $501 \times 401$ single-planetesimal integrations. For these simulations, we assume the temperature of the gas disc to be 200~K at 1~au. The disc evolution that was used follows that outlined in Appendix~\ref{sec:appendix_discmodel} and the pebble flux is computed using equation~\ref{eq:mdotf}. We also vary the timing of planetesimal formation $t_{\rm off}$. The top row of Fig.~\ref{fig:pebacc_contour} has $t_{\rm off} = 0.1~{\rm Myr}$, the middle row has $t_{\rm off} = 0.5~{\rm Myr}$ and the bottom row has $t_{\rm off} = 1~{\rm Myr}$. These three different starting times account for the delay in the timing of planetesimal formation in the disc. The left column has pebble fragmentation effects included while the right column does not.

It is clear from Fig.~\ref{fig:pebacc_contour} that the mass distribution is bimodal: planetesimals either grow large or they barely grow at all. This bimodal outcome is mainly dependent on the initial diameter of the planetesimals, which in turn determines their pebble accretion efficiency. If their diameter is below a critical value (which has a slight dependence on disc parameters), then they are in the geometric regime and thus accreting pebbles is very inefficient regardless of their distance to the Sun (purple region in the left side of all the panels in Fig.~\ref{fig:pebacc_contour}). If they are larger than the critical diameter, then they are in the settling regime wherein the pebble accretion efficiency is high, and will thus experience more growth. The planetesimal's critical diameter between the geometric and settling regimes depends on the location in the disc, the formation time of the planetesimal, whether pebble fragmentation at the snow line is assumed, and also very likely the temperature of the disc (although this factor is not investigated here directly).

It would appear that a small planetesimal $(M_{\rm i} = 10^{-5}~M_{\rm E}~{\rm for~example})$ attained a larger final mass compared to a large planetesimal $(M_{\rm i} = 10^{-3}~M_{\rm E}~{\rm for~example})$ located at the same semi-major axis, but this is actually not the case: we plot the mass ratio instead of the final mass. Both planetesimals reach the pebble isolation mass at the end of the simulations -- which is of the order of a few Earth masses (with a weak distance dependence). As such, a planetesimal with $M_{\rm i} = 10^{-5}~M_{\rm E}$ growing to $1~M_{\rm E}$ increased its mass by a factor of $\log(1/10^{-5}) = 5$ while the other planetesimal with $M_{\rm i} = 10^{-3}~M_{\rm E}$ increased its mass by a factor of $\log(1/10^{-3}) = 3$.

Without fragmentation (right column of Fig.~\ref{fig:pebacc_contour}), only planetesimals with $M \gtrsim 10^{-5}~M_{\rm E}~(D \gtrsim 580~{\rm km})$ will efficiently accrete pebbles, with a slight dependence on their initial location in the disc (the critical value for planetesimal diameter becomes smaller as we move closer to the Sun); everything smaller will barely accrete anything. Fragmentation (left column of Fig.~\ref{fig:pebacc_contour}), on the other hand, allows for the creation of a narrow annulus of large bodies between $\sim0.8~{\rm au}$ to $\sim1.2~{\rm au}$ whose exact extent depends on the formation time of the planetesimals relative to the evolution of the snow line. In our model the snow line does not reach inside 0.75~au so accretion inside of this distance can be inefficient if planetesimals are predominantly small $(M \lesssim 10^{-7}~M_{\rm E}~{\rm or}~D \lesssim 300~{\rm km})$ and formed late $(t_{\rm off} \gtrsim 1~{\rm Myr})$. In our fiducial model the snow line is at 1.2~au when $t+t_{\rm off}=1~{\rm Myr}$, and it passes 2~au at $t+t_{\rm off}= 0.44~{\rm Myr}$, thus substantial pebble accretion in the terrestrial planet region can only occur if planetesimals form very early, say within 0.5~Myr of the formation of the first solids \citep[e.g.][]{Kruijeretal2014}, and/or form large. In the asteroid belt region accretion generally stops really early on, at a time comparable to the formation age of the iron meteorites, and possibly before most chondrule formation ages -- typically about 1.5~Myr \citep{luuetal2015} -- because of the rapid inward migration of the snow line. Beyond 2~au accretion is generally insubstantial for small planetesimals.

The contour plot, although instructive, only provides information on the evolution of a single planetesimal. To understand the effects of pebble accretion in the early inner Solar System which likely hosted many planetesimals with a range of diameters, it is therefore necessary to turn to {\it N}-body simulations.
%%%%%%%%%%%%%%%%%%%%%%%%%%%%%%%%%%%

%%%%%%%%%%%%%%%%%%%%%%%%%%%%%%%%%%%
\section{Methods}
\label{sec:pebacc_methods}

\subsection{Gas disc model}
In our simulations the gas disc prescription is based on the model of \citet{Idaetal2016}, with the detailed equations given in Appendix~\ref{sec:appendix_discmodel}. The gas disc is assumed to be a steady-state accretion disc with a constant value for the accretion viscosity parameter $\alpha$ \citep{ShakuraSunyaev1973}. The temperature of the disc depends on two heating sources: viscous dissipation or stellar irradiation, which dominate in different regions of the disc. Equations describing the disc temperature profiles (equation~\ref{eq:T_visirr}) due to the aforementioned heating sources are obtained from empirical fits to the disc model by \citet{GaraudLin2007}. The thermal structure of the disc in turn determines the scale height and surface density. In our simulations, the gas disc dissipates away with time and was photoevaporated away in 500~kyr when the stellar accretion rate $\dot{M}_* < 10^{-9}~M_{\odot}~{\rm yr}^{-1}$.

In our simulations, the initial stellar accretion rate $\dot{M}_* = 2.63\times10^{-8}~M_{\odot}~{\rm yr}^{-1}$ and the initial pebble flux $\dot{M}_{\rm F} = 10^{-4}~M_{\rm E}~\rm{yr}^{-1}$. This corresponds to a disc age of $t_0 = 0.5~{\rm Myr}$. This value of $t_0$ in turn determines the subsequent values of $\dot{M}_*(t)$ and $\dot{M}_{\rm F}(t)$ throughout the simulation.  
%%%%%%%%%%%%%%%%%%%%%%%%%%%%%%%%%%%

%%%%%%%%%%%%%%%%%%%%%%%%%%%%%%%%%%%
\subsection{Pebble accretion}
The pebbles are assumed to have originated in the outer regions of the protoplanetary disc where most of the solids (dust) are located \citep[e.g.][]{YoudinShu2002,Garaud2007,Birnstieletal2012}. In this region, sub-micron-sized dust grains can grow into pebbles because their growth timescale is much shorter than their migration timescale and the collision rate is low \citep{Birnstieletal2012}. Upon reaching a critical diameter (or a critical Stokes number), the migration timescale becomes comparable to the growth timescale and the pebbles commence their inward migration towards the Sun by virtue of gas drag \citep{Idaetal2016} while their growth practically ceases. Due to the strong radial dependence of the growth timescale, there will be a location in the disc where dust clumps have just reached pebble size and start to migrate inwards. This is known as the pebble formation front and it moves outwards with time until it reaches the outer edge of the protoplanetary disc \citep{LambrechtsJohansen2014,Idaetal2016}, after which the pebble flux is severely reduced because all the solids in the disc have been consumed \citep[e.g][]{Chambers2016,Satoetal2016}.

We implemented the pebble accretion prescriptions of \cite{Idaetal2016} into the {\it N}-body code SyMBA \citep{Duncanetal1998} based on the method presented in \citet{Matsumuraetal2017}. Unlike \citet{Levisonetal2015,Levisonetal2015giantpl} we do not directly compute the growth of planetesimals by accretion of physical pebbles, but instead compute their mass increase based on the pebble flux at their respective locations in the disc. Pebble accretion onto planetesimals occurs outside in, that is, planetesimals at the outer edge of our solid disc are the first to encounter the pebbles. They will accrete a fraction of the pebbles $\dot{M}_{\rm p}$, reducing the pebble flux to $1-\dot{M}_{\rm p}/\dot{M}_{\rm F}$. The planetesimals that are next in line will see the reduced pebble flux and the amount of pebbles they can accrete is computed from the reduced pebble flux. Furthermore, when a planetesimal (or planetary embryo) reaches its pebble isolation mass $M_{\rm p,iso} \sim 1/2(H_{\rm g}/r)^3 M_*$ we assume that its accretion stops and pebbles are prevented from flowing past its orbit to other planetary bodies residing closer than it is to the Sun. At the snow line, we assume that pebbles were fragmented into grains of 1~mm on their path towards the Sun \citep{Morbidellietal2016}. We also study the outcome of simulations excluding this effect.
%%%%%%%%%%%%%%%%%%%%%%%%%%%%%%%%%%%

%%%%%%%%%%%%%%%%%%%%%%%%%%%%%%%%%%%
\subsection{Initial conditions for the N-body simulations}
\begin{figure*}
    \centering
	\includegraphics[width=0.8\textwidth]{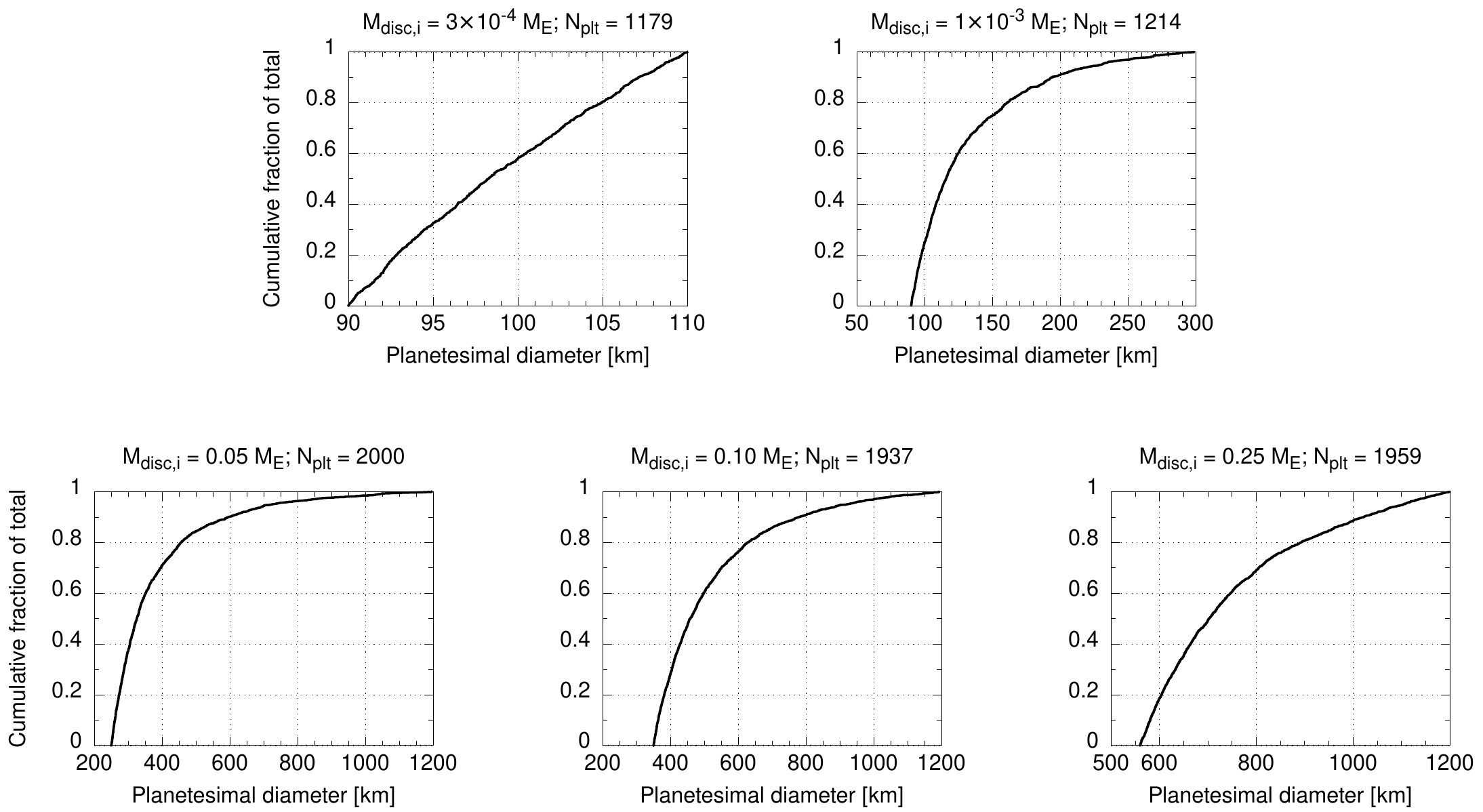}
    \caption{Cumulative distribution of planetesimal diameter $D$ for various values of initial solid disc mass $M_{\rm disc,i}$.}
    \label{fig:pebacc_initcond_D}
\end{figure*}

\begin{table}
	\centering
	\caption{Initial solid disc mass $M_{\rm disc,i}$ in units of Earth masses, range of planetesimal diameters $D$ in each disc, and initial number of planetesimals in each disc $N_{\rm plt}$ for the pebble accretion simulations. We repeat the simulations using the same set of initial conditions shown here for different values of initial disc temperature $T_{\rm 1~au} = 200~{\rm K},~250~{\rm K,~and}~300~{\rm K}$.}
    \label{tab:pebacc_initcond}
    \begin{tabular}{AAA}
    \toprule
    $M_{\rm disc,i}~(M_{\rm E})$ & $D~({\rm km})$ & $N_{\rm plt}$ \\ 
    \midrule
    $3\times10^{-4}$ & $90 - 110$   & $\sim 1200$ \\
    $1\times10^{-3}$ & $90 - 300$   & $\sim 1200$ \\
    0.05             & $250 - 1200$ & $\sim 2000$ \\
    0.10             & $350 - 1200$ & $\sim 2000$ \\
    0.25             & $560 - 1200$ & $\sim 2000$ \\
    \bottomrule
    \end{tabular}
\end{table}

Our planetesimal disc contains planetary bodies with an initial density of $\rho = 2500~{\rm kg~m}^{-3}$ and diameter $D$ assigned following a cumulative size-frequency distribution of $N(>D) \propto D^{-5/2}$ (Fig.~\ref{fig:pebacc_initcond_D}). Their orbital eccentricities $e$ and orbital inclinations $I$ are assigned at random from a uniform distribution with intervals $[0,0.01)$ and $[0^{\circ},1{^\circ})$, respectively. The remaining orbital angles (longitude of ascending node $\Omega$, argument of periapsis $\omega$ and mean anomaly $\mathcal{M}$) are assigned values between $0^{\circ}$ and $360^{\circ}$, also randomly chosen from a uniform distribution. The semi-major axis of the planetesimals ranges from $0.5~{\rm au} < a < 2.0~{\rm au}$ and they are distributed according to an $r^{-1}$ surface density profile. Our choice to limit the outer edge of the solid disc to 2~au is guided by the analytical results in Fig.~\ref{fig:pebacc_contour} and our intent to save computation time (our simulations contain a fairly large number of planetesimals that are all self-gravitating). In Fig.~\ref{fig:pebacc_contour} we showed that planetesimals smaller than $D \lesssim 500~{\rm km}$ located beyond 2~au are not expected to grow by much but smaller planetesimals could if pebble fragmentation is in force. Therefore we chose not to place any planetesimals beyond 2~au in the calculations and focus on the terrestrial planet region. Bodies with a diameter larger than 500~km can experience substantial growth, as was shown by \citet{BrasserMojzsis2020}.

The variables in our simulations are: (1) the initial mass of the solid disc $M_{\rm disc,i}$, (2) the range of planetesimal diameters $D$, and (3) the initial disc temperature at 1~au $(T_{\rm 1~au} =$~200~{\rm K},~250~{\rm K},~{\rm and}~300~{\rm K}) which corresponds to different disc scale heights $h_{\rm g}$ and pebble scale heights $h_{\rm peb}$. The lower mass discs ($M_{\rm disc,i} = 3\times10^{-4}~M_{\rm E},~1\times10^{-3}~M_{\rm E}$) contain $\sim 1200$ planetesimals while the more massive discs contain up to $\sim 2000$ planetesimals. All the planetesimals in our simulations are self-gravitating, that is, they are able to interact with each other via gravity. We summarise the initial conditions in Table~\ref{tab:pebacc_initcond}.

Our choice of planetesimal diameters are based on the results of previous works. The characteristic diameter of early planetesimals that formed via turbulent concentration or streaming instability in a Solar-like protoplanetary disc ranges from $\sim 100~{\rm km}$ up to $\sim 1000~{\rm km}$ \citep[e.g.,][]{Chambers2010,Johansenetal2014,KlahrSchreiber2020}, with a size-frequency distribution that follows a shallow power law $N(>D)\propto D^{-1.8}$ for diameter, corresponding to $N(>M) \propto M^{-0.6}$ for mass \citep[e.g.][]{Johansenetal2015,Simonetal2016,Schaeferetal2017,Abodetal2019}. Models of the collisional evolution of the asteroid belt also suggest that the primordial asteroids were typically 100~km in diameter \citep{Morbidellietal2009}. The combined results of these studies suggest that the initial planetesimal population likely consisted of many small planetesimals and a few massive bodies. Asteroid Vesta could be one of the early-formed massive bodies at the tail end of the distribution.

We opted for a different range of planetesimal diameters for different values of initial disc mass because we wanted to study with high enough resolution how pebble accretion affects both small and large planetesimals. If we were to adopt the same range of planetesimal diameters $(100~{\rm km} \leq D \leq 1200~{\rm km})$ for all disc masses, then we would have $N_{\rm plt} = 10-50$ for $M_{\rm disc,i} = 10^{-4}~M_{\rm E}$, $N_{\rm plt} = 100-400$ for $M_{\rm disc,i} = 10^{-3}~M_{\rm E}$, $N_{\rm plt} = 2500-3200$ for $M_{\rm disc,i} = 10^{-2}~M_{\rm E}$, $N_{\rm plt} \sim 14000$ for $M_{\rm disc,i} = 0.1~M_{\rm E}$, and $N_{\rm plt} \sim 31000$ for $M_{\rm disc,i} = 0.25~M_{\rm E}$. The resolution would be low for the low-mass discs and we could not really study the effect of pebble accretion onto a swarm of small planetesimals because any large planetesimal in the swarm would dominate the growth and potentially scatter the smaller planetesimals away. For the higher mass discs the computing time would take too long: according to our benchmarks of the code, integrating 14000 self-gravitating planetesimals for 4~Myr would take about a year on a 4.1~GHz 32-core AMD Threadripper CPU with OpenMP parallelisation assuming no loss through ejection or collision. For these reasons we chose to fix the maximum number of planetesimals for each disc and then shift the minimum and maximum diameter according to the disc mass. Nevertheless, we have run some simulations with the same planetesimal diameter range for all disc masses whose results are reported in Appendix~\ref{sec:appendix_lowmassdisc}.

In addition to the planetesimal disc, following the work of \citet{BrasserMojzsis2020} our initial set-up also includes a $0.01~M_{\rm E}$ planetary embryo placed at 5.6~au that would eventually become Jupiter. When Jupiter reaches its pebble isolation mass \citep[$20~M_{\rm E}$;][]{Lambrechtsetal2014} in about 1~Myr \citep{Kruijeretal2017jupiter}, it will open a partial gap in the disc around its orbit that prevents pebbles from beyond its orbit to spiral in towards the Sun, effectively shutting off the pebble flux to the inner Solar System \citep{Lambrechtsetal2014}, leaving only very small particles with sizes $\lesssim 100~\mu{\rm m}$ to pass through unobstructed \citep{PaardekooperMellema2006}.

We employ the SyMBA {\it N}-body code \citep{Duncanetal1998} modified to include pebble accretion, the effects of disc-induced orbital migration and orbital eccentricity damping \citep[Appendix~\ref{sec:appendix_migration}; e.g.][]{Tanakaetal2002,Matsumuraetal2017} and gas envelope accretion for massive bodies \citep[Appendix~\ref{sec:appendix_gasaccretion};][]{Ikomaetal2000,Matsumuraetal2017}. We also use the parallelised version from \citet{LauLee2021}. We also simulate the case where orbital migration is excluded to isolate the effects of the accretion of pebbles; these simulations are purely academic. We ran one simulation for each permutation of initial conditions (initial solid disc mass $M_{\rm disc,i}$, disc temperature $T_{\rm 1~au}$, fragmentation on/off, migration on/off). In total we ran 60 simulations, each for 4~Myr and with a time step of 0.01~yr.

We quantify the amount of mass increase in the inner Solar System resulting from the accretion of pebbles by computing the ratio of the final mass of the planetesimal disc versus its initial mass $M_{\rm disc,f}/M_{\rm disc,i}$. For the more massive discs $(M_{\rm disc,i} \geq 0.05~M_{\rm E})$, the larger planetesimals will accrete pebbles efficiently and they can grow to large masses quickly, as we will show in the following section. Therefore, the final disc mass for these cases do not only reflect growth via pebble accretion, but also additional growth due to mergers and gas envelope accretion. 
%%%%%%%%%%%%%%%%%%%%%%%%%%%%%%%%%%%

%%%%%%%%%%%%%%%%%%%%%%%%%%%%%%%%%%%
\section{Results}
\subsection{Amount of mass increase in the inner disc}
\begin{figure*}
    \centering
	\includegraphics[width=0.7\textwidth]{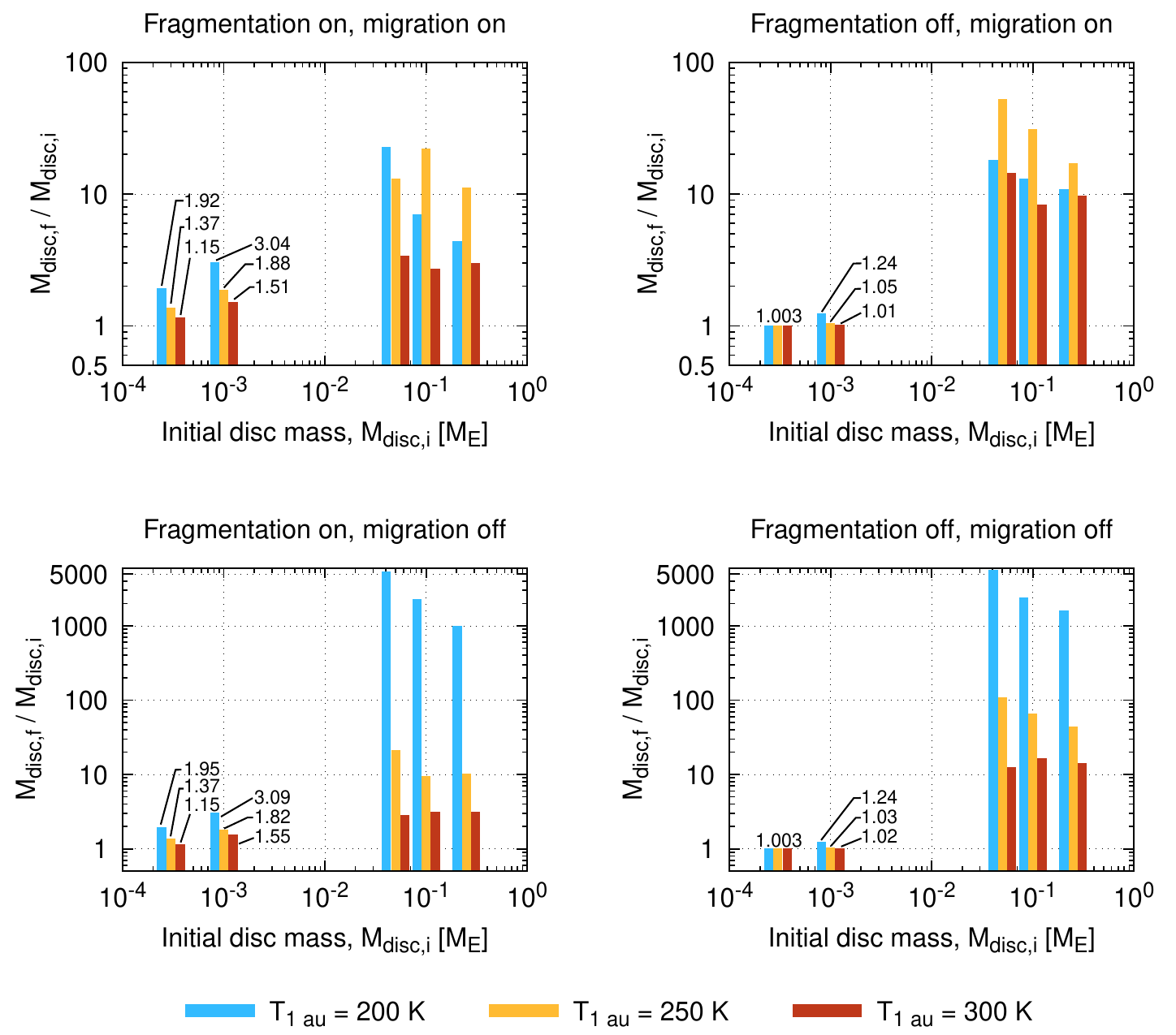}
    \caption{Ratio of final mass in the disc to its initial mass $(M_{\rm disc,f}/M_{\rm disc,i})$ in the region between 0.5~au and 3.0~au as a result of pebble accretion over a period of 4~Myr, plotted with respect to the initial disc mass and disc temperature. The accretion efficiency depends on the sizes of the planetesimals. Large amounts of mass in the form of pebbles can still be accreted by large planetesimals within 2~au before Jupiter reaches its pebble isolation mass.}
    \label{fig:pebacc_mfmi_vs_mdisc}
\end{figure*}

We show in Fig.~\ref{fig:pebacc_mfmi_vs_mdisc} the final mass in the disc between 0.5~au and 3.0~au as a fraction of its initial mass $(M_{\rm disc,f}/M_{\rm disc,i})$ for different disc temperatures at the end of the simulations $(t = 4~{\rm Myr})$. We find that the final mass of the solid disc depends mostly on its initial mass. A higher initial disc mass results in a relatively higher final disc mass because these discs have, on average, a higher number of large planetesimals which are more efficient at accreting pebbles. In discs containing mostly small planetesimals with $D \lesssim 300~{\rm km}~(M_{\rm disc,i} \leq 1\times10^{-3})$, the overall increase in mass is very limited as the planetesimals are not in the settling regime in which accretion proceeds efficiently. This outcome is similar to that of the single-planetesimal computations shown earlier in Fig.~\ref{fig:pebacc_contour}, which illustrates that a planetesimal's accretion regime abruptly transitions from the less efficient geometric regime to the more efficient settling regime when their masses (or diameters) are above a critical threshold given a specific set of gas disc parameters.

We also observe a general inverse correlation between the amount of mass increase and the initial temperature of the gas disc, i.e., more growth in cooler discs compared to warmer discs. This is because (1) for cooler discs the pebble scale height is lower and this increases the pebble accretion efficiency of the planetesimals \citep{Idaetal2016}, and (2) the snow line is on average closer to the Sun in cooler discs than in hotter discs. The combined effects of (1) and (2) results in a high accretion efficiency that allows planetesimals to grow to large sizes very quickly, with the extreme case forming gas giants in the asteroid belt when the disc is cool and migration is switched off. This trend with disc temperature does not hold for the cases with $M_{\rm disc,i} \geq 0.05~M_{\rm E}$ when migration is in force (top row of Fig.~\ref{fig:pebacc_mfmi_vs_mdisc}). In these cases, the cold discs appear to record less growth at the end of the simulations compared to warmer discs. The reason for this is that many of the large planetesimals that grow big early in the cold discs migrated towards the Sun and were removed from the simulation.

When comparing the top and bottom rows of Fig.~\ref{fig:pebacc_mfmi_vs_mdisc}, the effect of disc-induced orbital migration is very clear for solid discs containing large planetesimals. When migration effects are turned off as in the control cases (bottom row of Fig.~\ref{fig:pebacc_mfmi_vs_mdisc}), the trend of inverse correlation between mass increase and disc temperature reappears and cold discs record the most increase in disc mass reaching a few thousand times their initial values because some of the planetesimals in these discs attained masses high enough to trigger gas accretion \citep{Ikomaetal2000}. For the case of less massive discs $(M_{\rm disc,i} \leq 1\times10^{-3})$, orbital migration effects do not play a role in influencing the final disc mass because the planetesimals in these discs did not have the opportunity to grow to large sizes for orbital migration to kick in.

Pebble fragmentation at the snow line affects the final disc mass of less-massive and massive discs differently. When comparing the left and right columns of Fig.~\ref{fig:pebacc_mfmi_vs_mdisc}, we observe that when fragmentation effects are included the relative final mass in massive discs $(M_{\rm disc,i} \geq 0.05~M_{\rm E})$ is higher than in the low-mass discs with $M_{\rm disc,i} \leq 1\times10^{-3}~M_{\rm E}$. Fragmentation reduces the Stokes number of the pebbles and allows for planetesimals of smaller sizes to become more efficient at accreting pebbles, as discussed earlier in Section~\ref{sec:pebacc_fragmentation}. Without fragmentation, the discs containing planetesimals with $D\lesssim 110~{\rm km}$ barely grow at all (just a 0.3\% increase) while the discs with $M_{\rm disc,i} \leq 1\times10^{-3}~M_{\rm E}$ containing planetesimals with $D\lesssim 300~{\rm km}$ recorded a growth ranging from 1 -- 2\% for hot discs with migration to a maximum 24\% for cold discs, in contrast to a 15 -- 309\% mass increase when fragmentation is taken into account. These results thus confirm the analytical arguments that small planetesimals have difficulty accreting pebbles efficiently if the pebbles did not break into smaller sizes at the snow line. On the other hand, the lesser degree of growth for massive discs appears paradoxical, but this can be explained by the reduced overall accretion rate of the planetesimals when the Stokes number of the pebbles is reduced.

To summarise this section, our simulation results show that solid discs containing planetesimals larger than $\sim 300~{\rm km}$ in diameter increase their mass by at least a few times when they are exposed to an incoming pebble flux from the outer Solar System over the lifetime of the gas disc. As we have also included Jupiter in the simulations, our results also demonstrate that the more massive planetesimal discs in the inner Solar System are expected to have accreted at least a few times their initial mass in the form of pebbles before Jupiter reaches its isolation mass and shuts off the pebble flux. For lower-mass discs containing mostly planetesimals less than 300~km in diameter the mass increase is not as much as it is for higher-mass discs with more massive planetesimals because of their lower accretion efficiency. However, for the case of cold discs with fragmentation effects considered, the final mass in the disc can be as high as three times the initial mass.
%%%%%%%%%%%%%%%%%%%%%%%%%%%%%%%%%%%

%%%%%%%%%%%%%%%%%%%%%%%%%%%%%%%%%%%
\subsection{Distribution of solids in the disc}
\begin{figure}
    \centering
	\includegraphics[width=\columnwidth]{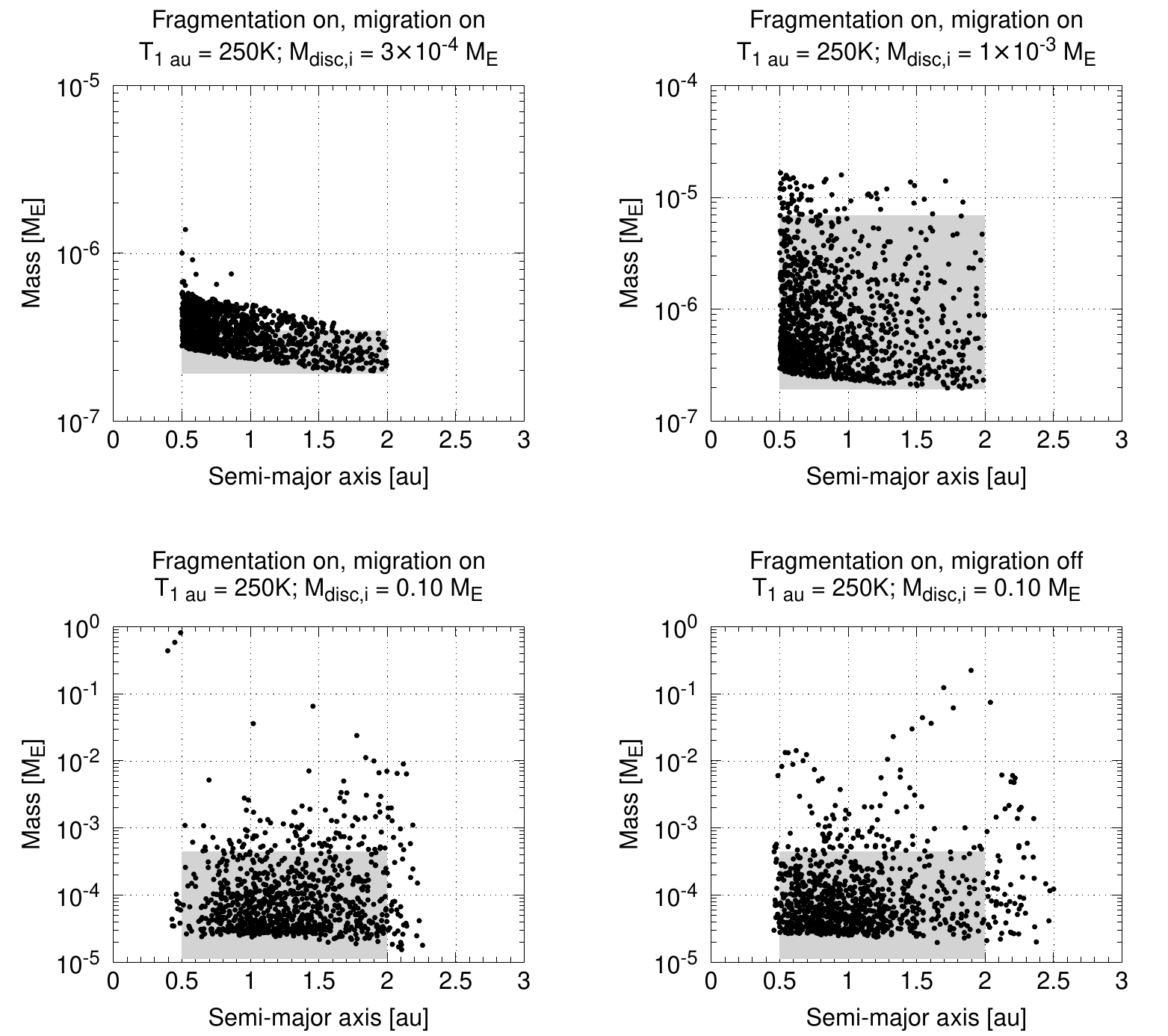}
    \caption{Distribution of mass in the inner disc at the end of the simulations. Shown here are the outcomes for different initial disc masses with the initial disc temperature at 1~au set to 250~K. The outcomes for discs with initial mass $M_{\rm disc,i} \leq 1\times10^{-3}~M_{\rm E}$ (top row) are representative of the simulations with different disc temperatures and regardless of whether migration and fragmentation are included. For the more massive discs we show the outcome for the case of $M_{\rm disc,i} = 0.10~M_{\rm E}$ (bottom row) when migration effects are included or excluded to illustrate its influence on the final mass distribution in the disc. The grey shaded region represents the initial mass and semi-major axis distribution of the planetesimals.}
    \label{fig:pebacc_m_vs_a}
\end{figure}

Having established that the amount of growth in the solid disc is dependent on the size (and mass) of its constituent planetesimals, we move on to report the distribution of solids at the end of the simulations for a few selected examples. We report selectively -- but representative of the results from all initial conditions -- the final mass distribution for warm discs $(T_{\rm 1~au} = 250~{\rm K})$ with fragmentation effects included (Fig.~\ref{fig:pebacc_m_vs_a}). For discs with initial mass $M_{\rm disc,i} \leq 1\times10^{-3}~M_{\rm E}$, the results are similar to the cases with fragmentation and/or migration excluded because (a) the planetesimals do not grow to sizes large enough to be affected by migration, and (b) the planetesimals do not grow much in the case of no fragmentation. Thus for these lower-mass discs, we can clearly see what is happening in the solid disc when the planetesimals are subjected to the pebble flux.

We observe a trend of inside-out growth: a higher mass increase for planetesimals located closer to the Sun compared to their counterparts located further away. This trend is clearest in the top-left panel of Fig.~\ref{fig:pebacc_m_vs_a} where we plot the result for the case where the solid disc contains only planetesimals with diameter less than $\sim 100~{\rm km}$. The minimum mass of the planetesimals does not have a dependence on the distance to the Sun initially (the lower bound of the grey shaded region is flat) but the lower bound of the final mass distribution (black dots) is higher when closer to the Sun but lower further away. This outcome is consistent with what was reported earlier in \citet{BrasserMojzsis2020}. For the case when $M_{\rm disc,i} = 1\times10^{-3}~M_{\rm E}$ (top-right panel of Fig.~\ref{fig:pebacc_m_vs_a}), the final mass distribution in the disc has a lower bound of $3\times10^{-7}~M_{\rm E}$ at 0.5~au while it is $2\times10^{-7}~M_{\rm E}$ at 2~au. The explanation for this inside-out growth outcome is the dependence of the pebble accretion rate on the semi-major axis \citep{Idaetal2016} -- the closer to the Sun, the higher the accretion rate. This outcome can also be understood using a more physical explanation: planetesimals in the region closer to the Sun see a higher concentration of pebbles because the pebbles are confined to an increasingly narrow annulus when they drift inwards, which is effectively an increased surface density in the disc, and thus it is easier to accrete pebbles in this region.

For the case of $M_{\rm disc,i} = 0.10~M_{\rm E}$ (bottom row of Fig.~\ref{fig:pebacc_m_vs_a}), the inside-out growth trend is masked because the planetesimals grow quickly due to their higher accretion efficiency and the inward migration of the snow line. In the example where migration is excluded (bottom-right panel of Fig.~\ref{fig:pebacc_m_vs_a}) we see two bumps, one each near the inner and outer edge of the solid disc, with the former attributed to inside-out growth and the latter caused by the passage of the snow line.

\begin{figure}
    \centering
	\includegraphics[width=\columnwidth]{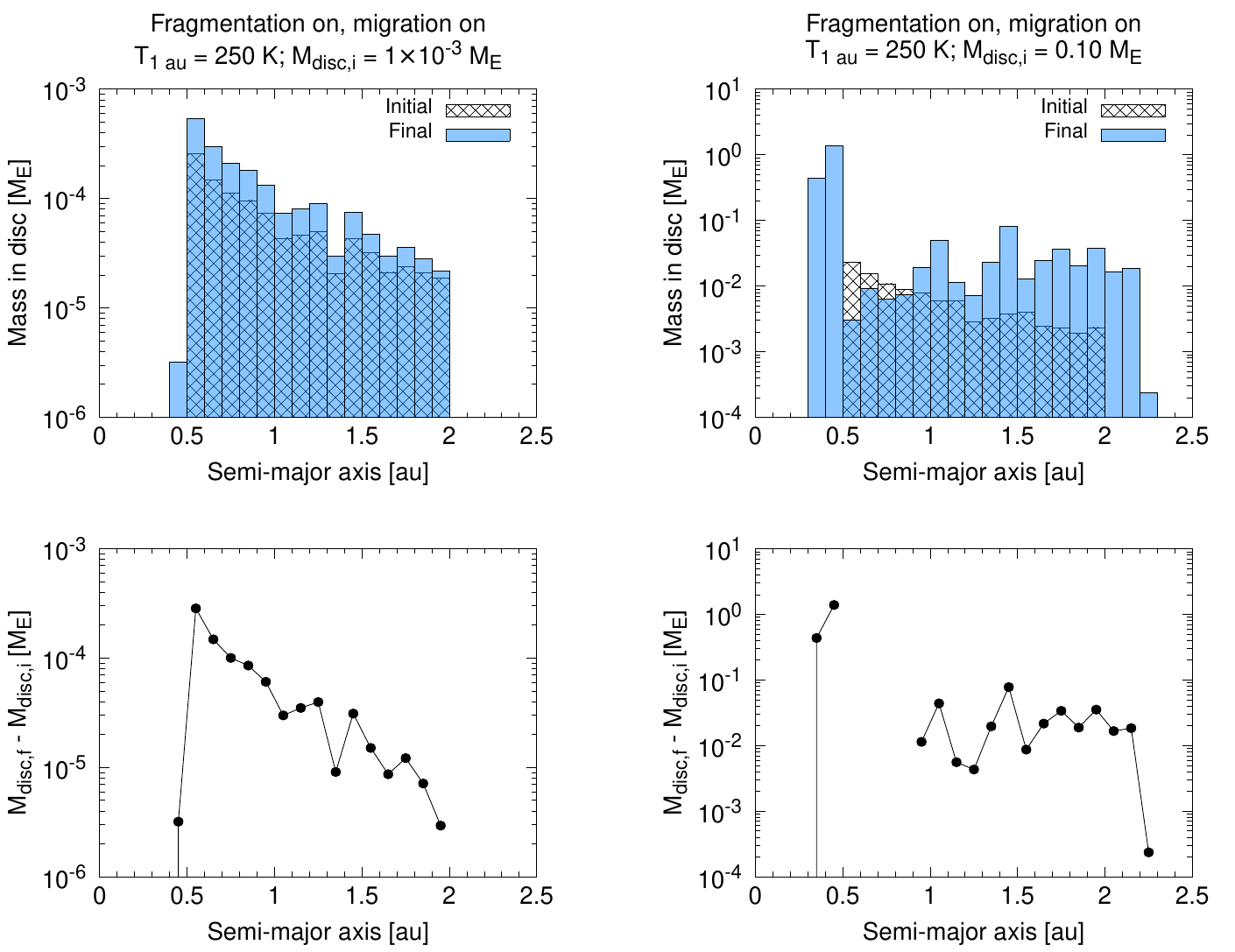}
    \caption{Initial and final mass distribution in each section of the disc (top row) and the corresponding amount of mass increase in each section (bottom row) for initial disc mass $1\times10^{-3}~M_{\rm E}$ and $0.10~M_{\rm E}$ and initial disc temperature of 250~K. The results shown in the left panel are representative of the outcomes for discs with low mass $(M_{\rm disc,i} \leq 1\times10^{-3}~M_{\rm E})$ while the right panel is representative of the general results for discs with high mass $(M_{\rm disc,i} \geq 0.05~M_{\rm E})$.}
    \label{fig:pebacc_mdiff}
\end{figure}

The data in Fig.~\ref{fig:pebacc_m_vs_a} can be presented as histograms which shows how different the final distribution is from the initial distribution. We show in the top row of Fig.~\ref{fig:pebacc_mdiff} the results for $M_{\rm disc,i} = 1\times10^{-3}~M_{\rm E}$ and $0.10~M_{\rm E}$ with migration to illustrate the two different possible outcomes depending on the size of planetesimals in the disc. In the less massive disc (left panel of Fig.~\ref{fig:pebacc_mdiff}), the planetesimals increase their mass but remain confined between 0.5 and 2.0~au. In the more massive disc (right panel of Fig.~\ref{fig:pebacc_mdiff}), the planetesimals are distributed in a wider region compared to their initial distribution because smaller planetesimals in the disc have their orbits perturbed by larger planetesimals and migration caused some of them to move inwards a little. If we measure the amount of mass increase in each semi-major axis bin (bottom row of Fig.~\ref{fig:pebacc_mdiff}), we find a smooth gradient in mass for the less massive disc, but not for the more massive disc.

We have shown that pebble accretion can generate a mass gradient in the terrestrial planet region. The mass gradient is only clearly seen in discs where the initial planetesimal diameters are less than $\sim 300~{\rm km}$; the gradient is not obvious if the disc contains many planetesimals larger than 300~km.
%%%%%%%%%%%%%%%%%%%%%%%%%%%%%%%%%%%

%%%%%%%%%%%%%%%%%%%%%%%%%%%%%%%%%%%
\section{Discussion}
\subsection{Jupiter’s capacity as a pebble barrier}
\begin{figure}
    \centering
	\includegraphics[width=\columnwidth]{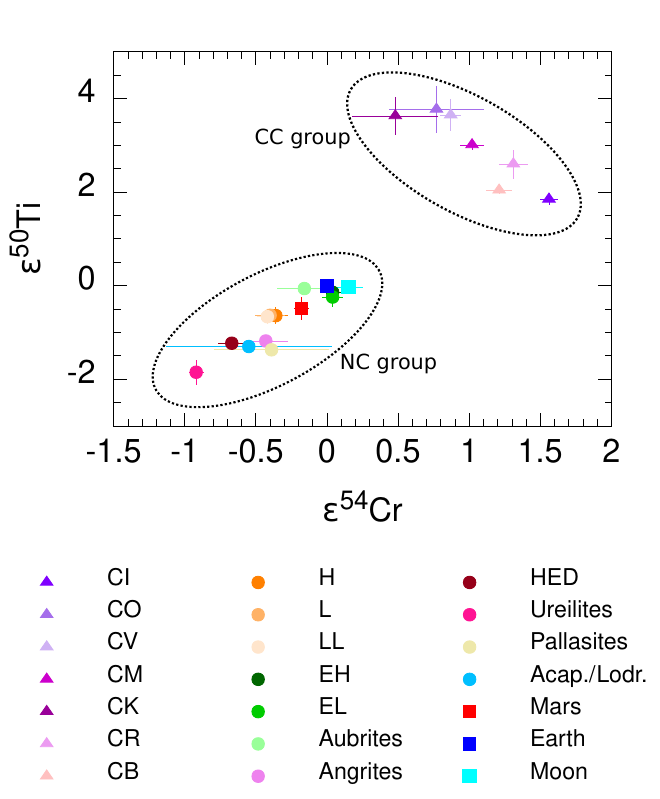}
    \caption{Measured $\varepsilon^{50}$Ti and $\varepsilon^{54}$Cr isotopic anomalies of known meteorites normalised to the values for the Earth. The meteorites fall into two distinct groups thought to represent the inner and outer Solar System, suggesting that the material in these two regions were kept separate at least until the formation of the chondrites. No data were available at the time of writing for $\varepsilon^{50}$Ti of iron meteorites. Acap. and Lodr. are abbreviations for Acapulcoites and Lodranites, respectively. C stands for carbonaceous and NC stands for non-carbonaceous meteorites. Data sourced from \citet{ShukolyukovLugmair2006,Trinquieretal2007,Trinquieretal2008,Trinquieretal2009,Qinetal2010a,Qinetal2010b,Yamashitaetal2010,Larsenetal2011,Petitatetal2011,Zhangetal2011,Zhangetal2012,YamakawaYin2014}.}
    \label{fig:isotopes}
\end{figure}

From the observed isotopic dichotomy among non-carbonaceous (NC) and carbonaceous (C) meteorites that sample the inner and outer Solar System, respectively (Fig.~\ref{fig:isotopes}), it is thought that the two reservoirs should have been separated very early and remain separated for at least a few Myr \citep{Kruijeretal2017jupiter}. Using measurements of molybdenum and tungsten isotopes \citet{Kruijeretal2017jupiter} demonstrated that iron meteorites also cluster into two distinct groups similar to the chondritic meteorites, suggesting that the two meteoritic reservoirs co-existed and remain separated over an extended period of time (at least $\sim 3$ to 4~Myr according to Kruijer et al.). The means to sustain the separation of the inner and outer Solar System has been suggested to be the growth of Jupiter \citep{Kruijeretal2017jupiter} which, upon attaining its pebble isolation mass, will create a gap in the disc and prevent solid material from beyond its orbit from entering the inner Solar System \citep{Lambrechtsetal2014}. However, there have been reports \citep{BrasserMojzsis2020,Kleineetal2020} challenging the efficiency of Jupiter as a competent barrier because a substantial amount of pebbles could have drifted into the inner Solar System before Jupiter reached its pebble isolation mass.

Previous works have shown that during the growth of Jupiter planetesimals located within its orbit can double their size \citep{BrasserMojzsis2020} or even attain the sizes of the current terrestrial planets \citep{Johansenetal2021}. Our {\it N}-body simulations with a larger number of planetesimals give a similar outcome: our planetesimal discs with initial mass $M_{\rm disc,i} \geq 0.05~M_{\rm E}$ and containing planetesimals with $D > 300~{\rm km}$ at least double their mass before Jupiter shuts off the pebble flux. This outcome holds for the initial disc temperatures we have tested (200 K, 250 K and 300 K), and is also independent of whether fragmentation effects are considered.

Given the substantial increase in mass in the inner Solar System contributed by pebbles despite Jupiter shutting off the pebble flux when it reaches its isolation mass, our results therefore indicate that it is unlikely for Jupiter to be the agent preventing the mixing of material from the two reservoirs. It should be more likely that some other factor(s) is/are at play at separating the inner and outer Solar System, for example, a structural gap in the protoplanetary disc located just within the orbit of Jupiter \citep{BrasserMojzsis2020}.
%%%%%%%%%%%%%%%%%%%%%%%%%%%%%%%%%%%

%%%%%%%%%%%%%%%%%%%%%%%%%%%%%%%%%%%
\subsection{Implications for cosmochemistry}
From mixing models, the maximum contribution of carbonaceous-like material to the composition of the Earth and Mars is capped at around 10\% \citep[e.g.][]{LoddersFegley1997,Sanloupetal1999,Lodders2000,TangDauphas2014,Fitoussietal2016,Dauphas2017,MahBrasser2021}. Based on nucleosynthetic anomalies of several isotopes such as $^{50}$Ti and $^{54}$Cr, the contribution of carbonaceous chondrites towards the making of the Earth is limited to 10 -- 25\% \citep[e.g.][]{Warren2011,Mezgeretal2020}. A mass contribution from outer Solar System material by amounts greater than 10\% poses problems for reconciling the difference in isotopic composition of the terrestrial planets with that of the outer Solar System, and a mass increase beyond 100\% will severely dilute the original composition of the inner disc and mostly homogenise it.

If we take 10\% as the upper limit for the amount of outer Solar System material that could be present in the Earth and Mars, then the successful cases from our numerical simulations are the cases which fulfil all of the following criteria:
\begin{itemize}
    \item $M_{\rm disc,i} \leq 1\times10^{-3}~M_{\rm E}$, or if the disc is more massive then all planetesimals have $D \leq 300~{\rm km}$, and
    \item $T_{\rm 1~au} \geq 250~{\rm K}$, and
    \item pebble fragmentation effects at the snow line are unimportant.
\end{itemize}
For these cases, the overall mass increase in the disc ranges from 0.3 -- 5\%, consistent with the constraint from mixing models and cosmochemistry data. The shortcoming for these cases, however, is that the total mass in the disc after 4~Myr is too little to build the terrestrial planets. To fulfil both requirements for the amount of mass increase $(\lesssim 10\%)$ and disc mass at the end of the pebble accretion phase $(\gtrsim 2~M_{\rm E})$, we should ideally start with a solid disc of $M_{\rm disc,i} = 1.8~M_{\rm E}$ which will grow to $2~M_{\rm E}$ assuming a 10\% mass increase and that the disc only contained planetesimals of diameter $D \lesssim 300~{\rm km}$ and that no mergers occurred during the stage of pebble accretion. However, modelling such a massive disc made up entirely of small planetesimals is at the limit of the capabilities of current hardware and {\it N}-body codes \citep{Wooetal2021}. We therefore reserve this study for future work. If the inner Solar System contained a high number of planetesimals with diameters 300~km and larger, we expect from our results that the isotopic composition of the planetesimals (and the planetary bodies that form from these planetesimals) would be akin to the outer Solar System -- inconsistent with cosmochemistry data.

Will the outcome of the {\it N}-body simulations change if we assume a changing composition of the pebbles? In our study, the implicit assumption is that the pebbles have the same composition, which we leave unspecified but one could assume any of the various carbonaceous chondrites as a proxy. In reality, the pebbles that formed at different times and at different locations in the disc could potentially have different compositions \citep{Lichtenbergetal2021}. The diversity in the isotopic compositions of the carbonaceous chondrites possibly reflects their distinct formation location and/or their formation time \citep{Deschetal2018}, although this is difficult to prove. By looking at the isotopes of O, Ti and Cr, the CI and CO chondrites plot in end-member positions of the carbonaceous meteorite group (see Fig.~\ref{fig:isotopes}), which could mean that there was a compositional gradient in the outer Solar System when these bodies formed \citep[e.g.][]{Trinquieretal2009,Mezgeretal2020}. To model the effect of time-dependent pebble compositions, we need to understand how the compositional gradient scales with heliocentric distance. This information is currently unavailable, although \citet{Deschetal2018} suggested that the CI chondrites formed farthest away from the Sun while CO chondrites formed closer to the Sun, but this suggestion is based mostly on model predictions.

\begin{figure}
    \centering
	\includegraphics[width=\columnwidth]{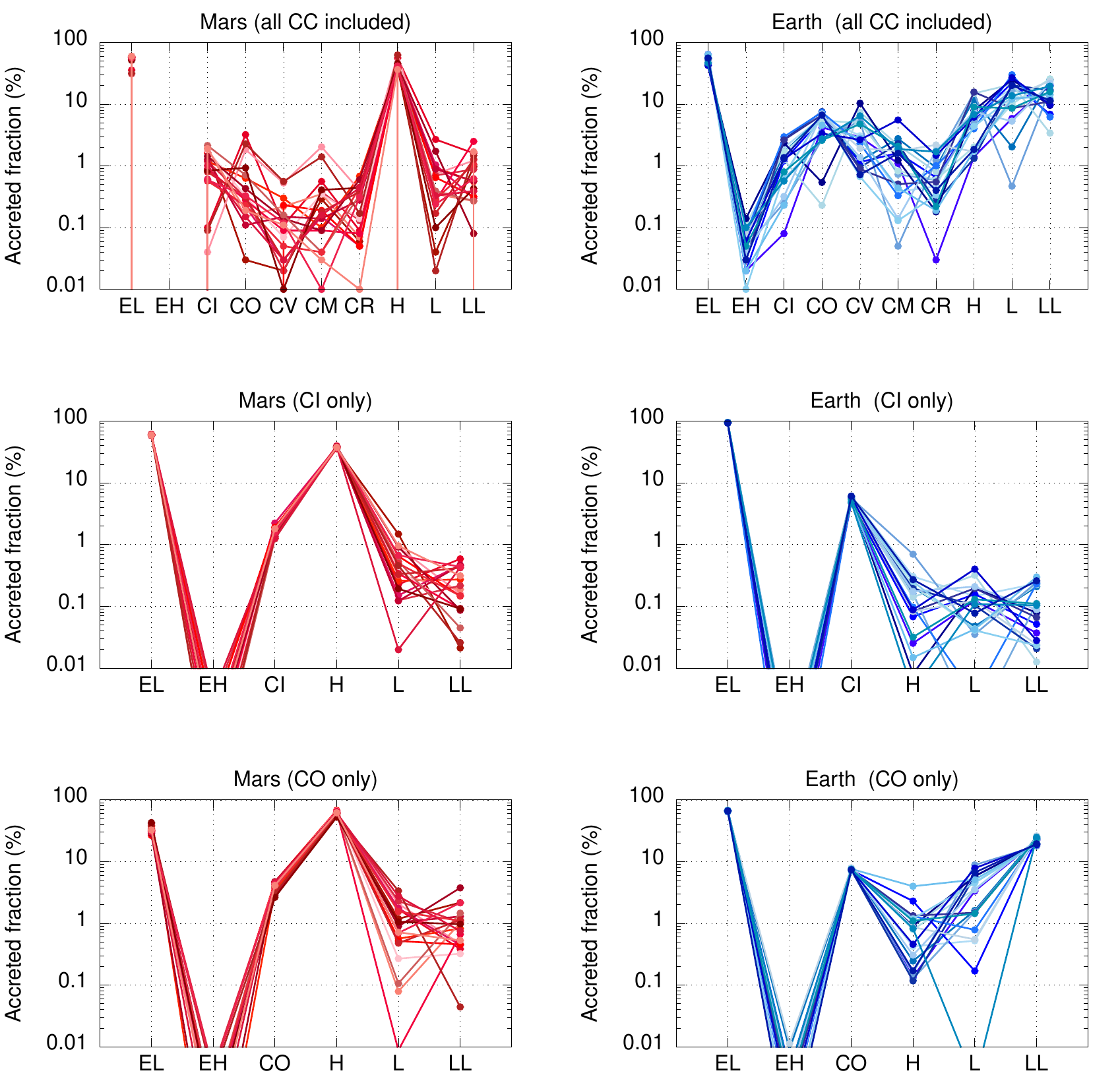}
    \caption{Comparison of Monte Carlo mixing model simulation outcomes. Top row shows the best-fit contributions of enstatite chondrites (EL, EH), ordinary chondrites (H, L, LL) and carbonaceous chondrites (CI, CO, CV, CM, CR) to the building blocks of Earth and Mars. The top row is the same plot that was presented in \citet{MahBrasser2021}. The middle and bottom rows show the results if we only include CI or CO chondrites (end-members of the carbonaceous chondrite group for O, Cr and Ti isotopes) in the model. All the results presented are computed from 20 Monte Carlo mixing model simulations based on the work of \citet{Dauphas2017}. The mixing model uses the isotopic anomalies in $\Delta^{17}$O, $\varepsilon^{50}$Ti, $\varepsilon^{54}$Cr, $\varepsilon^{62}$Ni and $\varepsilon^{92}$Mo for the meteorites and Earth and Mars as constraints. The contribution of CI or CO chondrites to the Earth and Mars is limited to $\leq 10\%$.}
    \label{fig:isocomp_mcmc}
\end{figure}

One possible way out of this impasse is to do a similar end-member study by estimating the contribution of CI or CO chondrites to the composition of the terrestrial planets using our isotope mixing models from \citet{MahBrasser2021} -- which is based on the model of \citet{Dauphas2017}. We show in Fig.~\ref{fig:isocomp_mcmc} the computed best-fit composition of the Earth and Mars as a combination of enstatite chondrites, ordinary chondrites and carbonaceous chondrites. The top panels are results previously reported in \citet{MahBrasser2021}. The middle and bottom rows include only CI or CO chondrites respectively. With only CI or CO chondrites, the mixing model returns the maximum contribution of these chondrite types as $\leq 10\%$ (Fig.~\ref{fig:isocomp_mcmc}), similar to the results obtained by including various other types of carbonaceous chondrites \citep[e.g.][]{Dauphas2017,MahBrasser2021}. For Earth and Mars the contribution from H-chondrites is most strongly affected by the choice of either CI or CO. Thus, even if we were to assume a time-dependent variation in the composition of the pebbles from CO-like at the beginning to CI-like at the end (or vice versa), it still remains difficult for our {\it N}-body simulation results to be reconciled with meteorite isotope data of the inner planets: the amount of jovian materials added to the terrestrial planets cannot be more than 10\%.

Let us consider the possible outcomes if we were to instead assume a specific composition for the planetesimals in the inner Solar System. We note that this is also speculative as it is not known in detail how the nucleosynthetic isotopes are distributed in the inner disc, although there are suggestions that they follow a heliocentric gradient \citep[e.g.][]{Trinquieretal2009,Yamakawaetal2010}. Nevertheless, we can reason that given the vastly distinct isotopic compositions of the NC and C meteorites in O, Ti and Cr for example, a large contribution of jovian material would require the terrestrial material to be initially very depleted in these isotopes so that the final outcome matches what is observed today for the sampled inner Solar System bodies. If the initial composition of the inner Solar System was, for example, ureilite-like then the accretion of pebbles with CI-like composition has been demonstrated to be consistent with the $\varepsilon^{48}$Ca compositions of Earth, Mars and asteroid Vesta \citep{Schilleretal2018}. However, the suggested combination of ureilite + CI (end-members of the NC and C groups) in the same proportion that works for Ca does not hold for other isotopes such as O, Ti and Cr.

The works of \citet{Schilleretal2018,Schilleretal2020} argue for pebble accretion in the inner Solar System based on $\varepsilon^{48}$Ca and $\varepsilon^{54}$Fe isotopes in the Earth and other meteorites. Their combined results argue for the Earth having 40\% CI-chondrite-like material in its mantle. For the ureilite parent body \citet{Schilleretal2018} list $\varepsilon^{48}{\rm Ca} = -1.46\pm0.46$ and for the CI chondrite parent body $\varepsilon^{48}{\rm Ca} = 2.06\pm0.085$. If we denote by $x$ the fraction of CI chondrite in Earth's mantle, then according to \citet{Schilleretal2018} $x$ is the solution to $2.06x-1.46(1-x) = 0$, which yields $x = 0.41\pm0.12~(2\sigma)$; the uncertainties were computed using a Monte Carlo method. If these results are correct, they should also hold for the isotopes of other elements commonly used as tracers, such as the multiple oxygen isotopes, and nucleosynthetic tracers such as $^{50}$Ti and $^{54}$Cr because the Earth can only be made up of a single mixture of sources rather than different mixtures for different elements. For the ureilite parent body $\Delta^{17}{\rm O} = -1.16\pm0.55$ \citep{ClaytonMayeda1996} and $\Delta^{17}{\rm O} = 0.39\pm0.14$ for the CI chondrites \citep{ClaytonMayeda1996}. The required mass fraction of CI chondrite based on oxygen isotopes is $x = 0.75\pm0.23~(2\sigma)$. This fraction deviates significantly compared to the fraction advocated by \citet{Schilleretal2018} based on calcium isotopes (i.e. 40\%). In addition, if molybdenum isotopes are considered then there is no acceptable solution for $x$ as the isotopic anomalies in $\varepsilon^{92}$Mo for both ureilites and CI chondrites have deficits relative to the Earth. We arrive at the same conclusion as well if we were to consider the more abundant isotopes of molybdenum $(\varepsilon^{94,95}{\rm Mo})$ in excess in CI and ureilites compared to Earth.

The above analysis shows that it is difficult to establish whether or not pebble accretion occurred in the inner Solar System when using isotopes alone because the end-member case of ureilite + CI chondrite could be invoked as a mixture to explain the isotopic anomalies of the terrestrial planets (albeit not always in the same proportions). Indeed, by looking at single isotopes independently it is always possible to find a combination of end-member meteoritic reservoirs with the correct mixing proportion to explain the isotopic composition of a planetary body. The difficulty is to find a single mixing ratio that matches all isotopes simultaneously as was done by \citet{Fitoussietal2016} and by \citet{Dauphas2017}. It is therefore necessary to look beyond nucleosynthetic isotopes alone to establish whether the simple model suggested by \citet{Schilleretal2018,Schilleretal2020} and pebble accretion can account for the growth and composition of the terrestrial planets. In the following, we examine the constraints from the major elemental abundances in known meteoritic reservoirs and planetary bodies.

\begin{figure}
    \centering
	\includegraphics[width=\columnwidth]{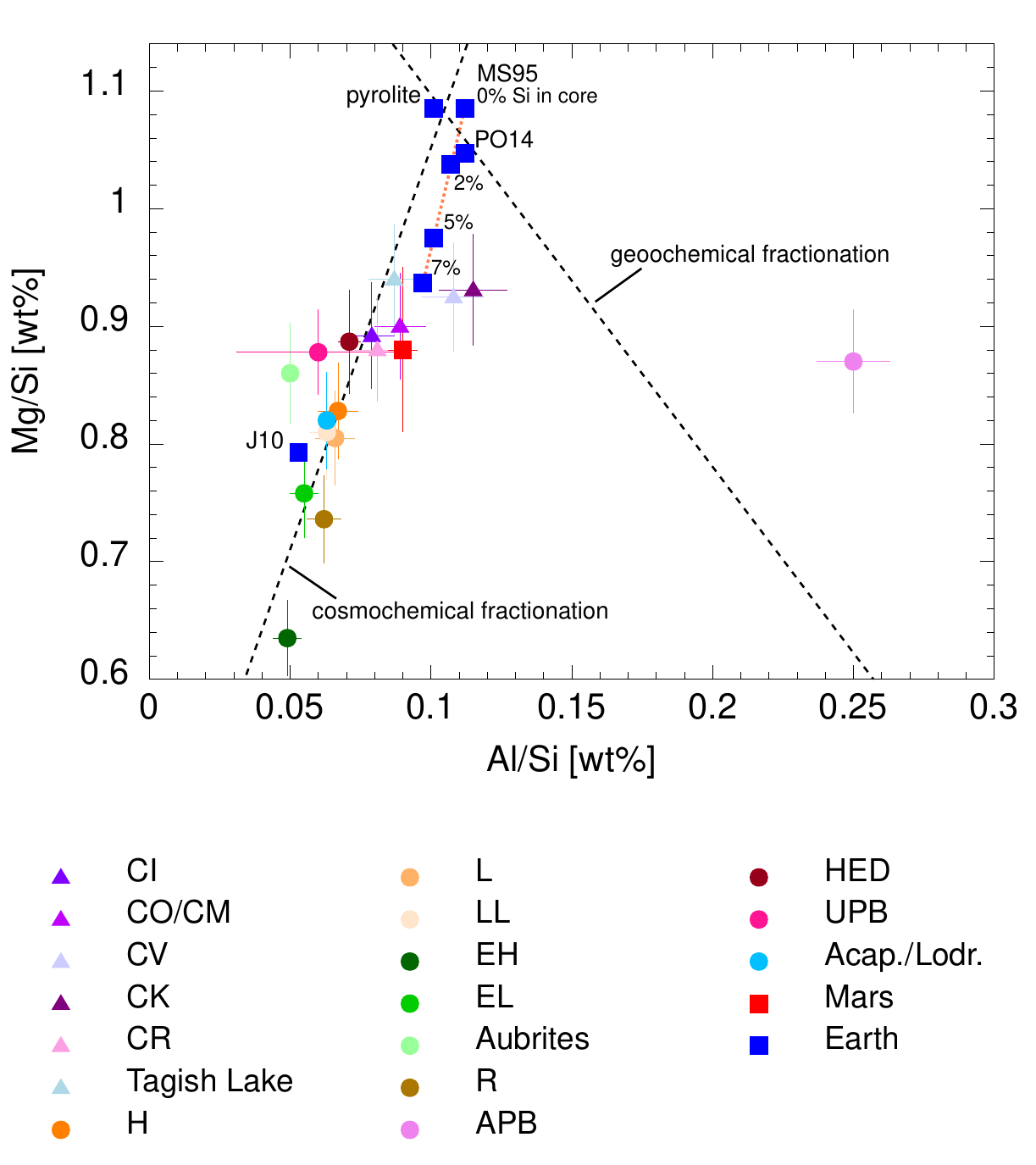}
    \caption{Mg/Si versus Al/Si for various chondrites, achondrites, Mars and the Earth. For the Earth, we also show the elemental abundance ratios of the Earth for different amounts of Si in the core. The fractionation lines and pyrolite composition are from \citet{Jagoutzetal1979}. Error bars are computed using the average variations for Mg/Si (5\%) and Al/Si (10\%) determinations for each parent body. APB = angrite parent body; UPB = ureilite parent body. Elemental abundance ratio for the Earth sourced from \citet{McDonoughSun1995,Javoyetal2010,PalmeONeill2014} while the data for meteorites are from \citet{MasonWiik1962,Vonmichaelisetal1968,Ahrensetal1973,ConsolmagnoDrake1977,Dreibusetal1977,Hertogenetal1977,Morganetal1978,Jagoutzetal1979,WattersPrinz1979,Jarosewichetal1987,WassonKallemeyn1988,Jarosewich1990,Kallemeynetal1991,Kallemeynetal1994,Kallemeynetal1996,Kongetal1997,Mittlefehldtetal1998,Goodrich1999,Longhi1999,Brownetal2000,WolfPalme2001,Greenwoodetal2010,Bischoffetal2011,Strackeetal2012,Blinovaetal2014,Hewinsetal2014,Palmeetal2014,CollinetGrove2020,YoshizakiMcDonough2020}.}
    \label{fig:elements}
\end{figure}

From the perspective of the major elemental abundances, the combination of chondritic (and perhaps achondritic) meteorites has been reported to be incapable of reproducing the Mg/Si ratio of the terrestrial planets \citep[e.g.][]{DrakeRighter2002,Dauphasetal2015MgSi} because the Earth has an end-member composition that is unlike any of the known meteorites. Following \citet{DrakeRighter2002} we show in Fig.~\ref{fig:elements} the Mg/Si versus Al/Si elemental ratios for the terrestrial planets and the known meteorite parent body compositions. The CI chondrites are similar in elemental composition to the Sun's photosphere \citep[e.g.][]{AndersGrevesse1989,Lodders2003} and there is little variation amongst the carbonaceous chondrites. The non-carbonaceous meteorites, which formed in the same reservoir as the Earth and Mars, all have lower Mg/Si ratios due to nebular and planetary processes \citep{Larimer1979,Alexander2019}; they all have more or less forsterite- or enstatite-rich compositions \citep{Dauphasetal2015MgSi}.

Possible compositions of the bulk Silicate Earth (BSE) computed from models based on terrestrial rocks and chondrites \citep[e.g.][]{McDonoughSun1995,PalmeONeill2014} place it at an end-member location in Fig.~\ref{fig:elements} with an Mg/Si ratio higher than the known meteorites. An exception is the result of \citet{Javoyetal2010} (blue square labelled with J10 in the figure) which is obtained by assuming that the Earth is made up entirely of enstatite chondrites. This result should however be interpreted with caution as a more recent study examining the $\delta^{30}$Si and Mg/Si composition of the Earth showed that it is similar to the enstatite chondrites isotopically but not chemically \citep{Dauphasetal2015MgSi}.

However, the exact composition of the BSE is in fact uncertain. This is partly due to uncertainties in the composition of the lower mantle and the amount of Si in the core. The models of \citet{McDonoughSun1995} and \citet{PalmeONeill2014} assume a homogeneous composition for the whole mantle (pyrolitic, with Mg/Si $\sim1.3$) while the model of \citet{Javoyetal2010} requires that the upper and lower mantle differ in their compositions. Results from seismic tomography observations of subducted crustal slabs \citep{FukaoObayashi2013}, computations and measurements of mineral elasticities \citep{Wangetal2015,Kurnosovetal2018} support a compositionally homogeneous mantle, although there are alternative proposals for a non-pyrolitic and silicon-enriched lower mantle based on its sound-velocity structure \citep{Murakamietal2012,Mashinoetal2020}. Be that as it may, the Mg/Si ratios of the lower mantle derived from these studies ($\sim 1.0$ for \citealt{Murakamietal2012} and 1.14 for \citealt{Mashinoetal2020}) still lie above the Mg/Si ratio of the known meteorites.

The amount of Si fractionated into the Earth's core is estimated to be between 2 wt\% to 7 wt\% based on geochemical and geophysical constraints \citep[e.g.][]{Badroetal2007,Fitoussietal2009,Moynieretal2020}. The Earth's chemical composition is similar within uncertainties to the carbonaceous chondrites if more than 5 wt\% of the Si is in the core. Nevertheless, this still does not change the fact that the Earth's chemical composition cannot be reproduced by any mixture of known planetary materials, not even the combination of ureilites and CI chondrites as was suggested by \citet{Schilleretal2018,Schilleretal2020} because this mixture would lower the Mg/Si ratio and requires the Earth to have more than 7 wt\% Si in the core. Furthermore, the current meteorite repository is likely incomplete and it is therefore an incautious assumption to make that these represent all the possible compositions of the material in the protoplanetary disc. Given the Earth's end-member composition, it has been suggested that a major component of the Earth $(\sim 40\%)$ could originate from the most refractory part of the inner Solar System that we do not have samples from \citep{Morbidellietal2020,Frossardetal2021} and later mixed with less thermally-processed planetesimals such as the known achondrite and chondrite parent bodies (Fig.~\ref{fig:elements}).

\begin{figure}
    \centering
	\includegraphics[width=\columnwidth]{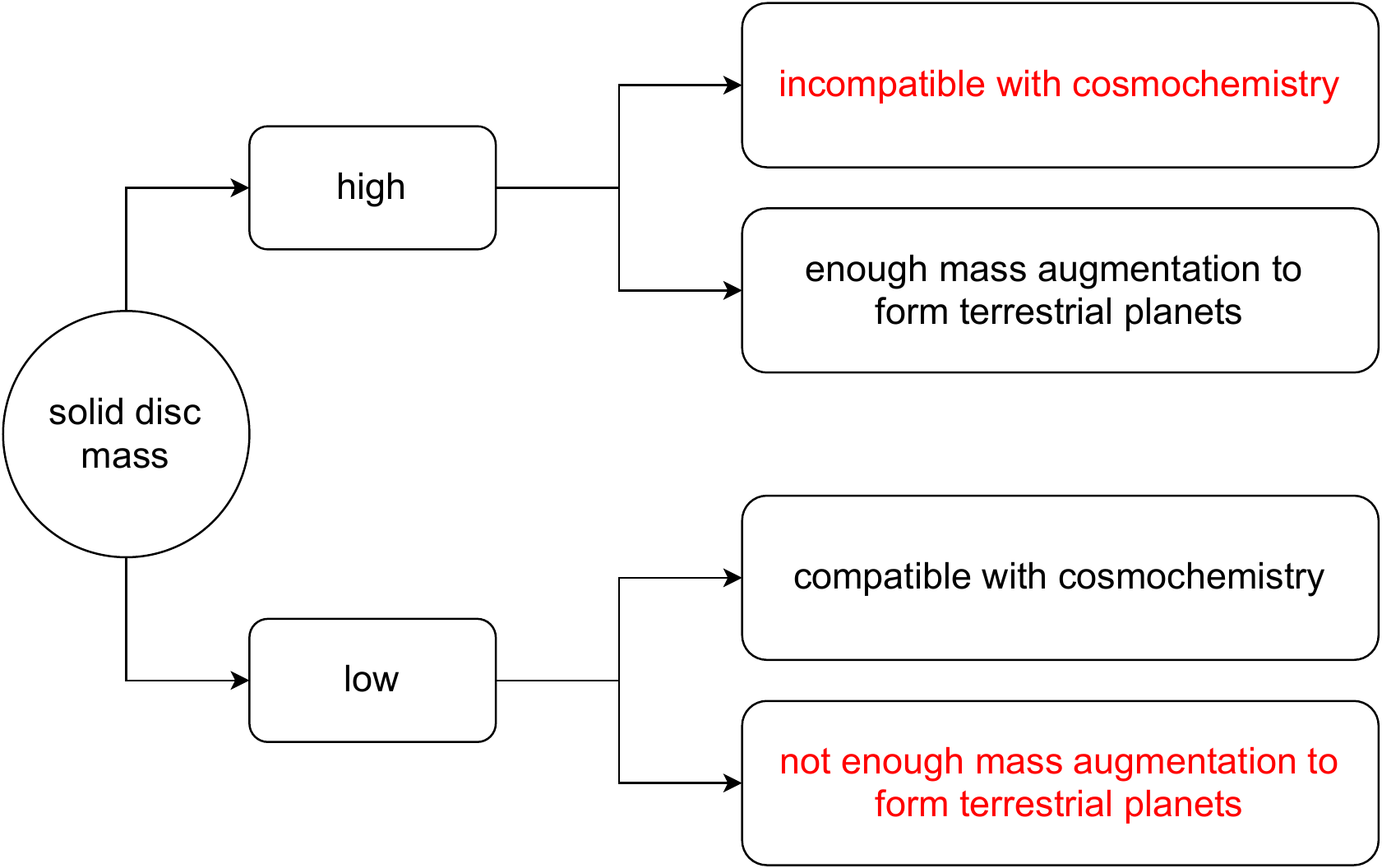}
    \caption{Flow chart summarising the two possible outcomes of applying the pebble accretion mechanism to the inner Solar System based on current understanding and without introducing additional ad-hoc assumptions. Our {\it N}-body simulation results show that regardless of the initial mass in the solid disc, there is no combination of initial conditions that can fulfil the constraints from both cosmochemistry and having sufficient disc mass to form the terrestrial planets.}
    \label{fig:flowchart}
\end{figure}

It seems unlikely that assuming the initial composition of the inner Solar System to be very depleted in the isotopes that the C group is enriched in (or vice versa) would alter the outcome of the numerical simulations. From the above discussion on the possibility of reconciling our simulation results with the constraints from cosmochemistry, the evidence appears to be against a strong contribution from pebble accretion in the inner Solar System. Fig.~\ref{fig:flowchart} summarises the main message we want to convey from this work. For pebble accretion as we currently understand it to work for the inner Solar System it would thus be necessary to introduce specific assumptions, such as those invoked by \citet{Levisonetal2015} and \citet{Johansenetal2021}. The need to call upon additional assumptions to make pebble accretion work in the inner Solar System in turn raises the question of whether it is absolutely necessary to use this model to explain terrestrial planet formation, for we already have on hand many other planet formation models which can reproduce the features of the inner Solar System by planetesimal mergers.
%%%%%%%%%%%%%%%%%%%%%%%%%%%%%%%%%%%

%%%%%%%%%%%%%%%%%%%%%%%%%%%%%%%%%%%
\section{Conclusions}
We report the outcome of {\it N}-body simulations investigating the growth of a disc of planetesimals with a range of diameters in the terrestrial planet region when they are subjected to a flux of pebbles originating from the outer Solar System. At the end of the simulations $(t = 4~{\rm Myr})$, we find that planetesimal discs made up of many planetesimals with diameters greater than 300 km will increase their mass by at least few times their initial mass, depending on the temperature of the gas disc and if gas-induced migration effects or pebble fragmentation effects at the snow line are included. This is despite Jupiter almost halting the pebble flux when it reaches its isolation mass. Such a large amount of mass increase contributed by outer Solar System pebbles would imply that the isotopic and chemical composition of the planetesimals will be replaced by the signatures of the pebbles. Planetary bodies formed by the subsequent collisions among these planetesimals are expected to have isotopic compositions similar to the outer Solar System bodies, which is inconsistent with the isotopic dichotomy revealed by currently available meteorite samples \citep{Warren2011}.

On the other hand, if the planetesimals in the inner Solar System are mostly bodies with diameters less than 300 km then the amount of mass increase in the disc is limited. The amount of mass increase ranges from 15\% to about 300\%, depending on the initial mass of the planetesimal disc and assuming that the pebbles fragment into smaller-sized grains at the snow line. The amount of mass accreted in the form of pebbles is, however, still considered to be quite high based on the results of mixing models that the maximum contribution of outer Solar System material to the mass of the Earth is about 10\% \citep{Dauphas2017}. In the case where we disregard the fragmentation effects, the mass increase in the disc can be less than 10\% the initial disc mass. However, the more important problem for these discs with small planetesimals is that the final mass in the disc by the time the gas disc dissipates is insufficient to form the terrestrial planets.

It is more likely that the solid disc in the inner Solar System was more massive than $10^{-3}~M_{\rm E}$ and thus should contain more planetesimals larger than 300~km in diameter. Since the growth of these planetesimals proceeds at a faster rate due to their higher accretion efficiency, it is expected that the contribution by pebbles to the mass in the terrestrial planet region will be very large. Even if we were to consider possible end-member compositions for the pebbles and the inner Solar System planetesimals, we still find it difficult for our simulation results to reconcile with cosmochemistry constraints. We therefore suggest that it is rather unlikely for pebble accretion as we currently understand it to play a major role in the formation of the terrestrial planets.

Compared to the work of \citet{Lambrechtsetal2019}, our results for the case of $M_{\rm disc,i} = 0.25~M_{\rm E}$ and without fragmentation (the only case for which a direct comparison can be made) are similar: the planetesimal disc recorded a more than ten times increase in its total mass during the lifetime of the gas disc. \citet{Lambrechtsetal2019} suggested that when Jupiter reached its pebble isolation mass it would stop the inflow of pebbles to the inner Solar System, but we show here that the mass in pebbles that entered the inner Solar System before it was shut off by the growing Jupiter already `contaminates' the inner Solar System with too much outer Solar System material. This comports with the results of \citet{BrasserMojzsis2020}. Our results are, however, at odds with the works of \citet{Levisonetal2015} and \citet{Johansenetal2021} mainly because of the different assumptions made. We assumed that the pebbles formed in the outer Solar System beyond the orbit of Jupiter whereas these previous studies assume for example, that (a) the pebbles form in the inner Solar System \citep{Levisonetal2015}, or that (b) there was a change in the isotopic composition of the pebbles from non-carbonaceous-chondrite-like to carbonaceous-chondrite-like within the lifetime of the gas disc \citep{Johansenetal2021}. As the pebble accretion mechanism is widely studied in the community, our understanding of it is gradually improving and this will provide opportunities for further examinations of the plausibility and validity of the assumptions used by current works.
%%%%%%%%%%%%%%%%%%%%%%%%%%%%%%%%%%%

%%%%%%%%%%%%%%%%%%%%%%%%%%%%%%%%%%%
\section*{Acknowledgements}

J.M. and R.B. thank Tommy Lau and Man Hoi Lee at the University of Hong Kong for their parallelised version of the SyMBA {\it N}-body code. J.M. acknowledges the support of the DFG priority program SPP 1992 ``Exploring the Diversity of Extrasolar Planets'' (BI 1880/3-1). R.B. and S.J.M. thank the Research Centre for Astronomy and Earth Sciences (Budapest, Hungary) for support of the Origins Research Institute (ORI). We thank the reviewer for comments which helped improve the clarity and flow of the manuscript.

%%%%%%%%%%%%%%%%%%%%%%%%%%%%%%%%%%%%%%%%%%%%%%%%%%
\section*{Data Availability}
The data underlying this article will be shared on reasonable request to the corresponding author.
 
%The inclusion of a Data Availability Statement is a requirement for articles published in MNRAS. Data Availability Statements provide a standardised format for readers to understand the availability of data underlying the research results described in the article. The statement may refer to original data generated in the course of the study or to third-party data analysed in the article. The statement should describe and provide means of access, where possible, by linking to the data or providing the required accession numbers for the relevant databases or DOIs.



\bibliographystyle{mnras}
\bibliography{pebble} % if your bibtex file is called example.bib

%\begin{thebibliography}{99}
%\end{thebibliography}

%%%%%%%%%%%%%%%%%%%%%%%%%%%%%%%%%%%%%%%%%%%%%%%%%%

%%%%%%%%%%%%%%%%% APPENDICES %%%%%%%%%%%%%%%%%%%%%
\appendix
\section{Gas disc model}
\label{sec:appendix_discmodel}
The gas disc model employed in our simulations is based on the model of \citet{Idaetal2016}. The accretion rate of the gas onto the central star (the Sun in our case) is assumed to occur at a steady rate given by
\begin{equation}
    \dot{M}_*=3\pi \alpha \Sigma_{\rm g} H_{\rm g}^2 \Omega_{\rm K},
\label{eq:dotmstar}
\end{equation}
where $\Sigma_{\rm g}$ is the gas surface density, $H_{\rm g}$ is the disc scale height and $\Omega_{\rm K}$ is the Keplerian frequency. The disc $\alpha$-viscosity parameter $\alpha$ was set to $10^{-3}$ and assumed to be constant throughout \citep{ShakuraSunyaev1973}. It is given by
\begin{equation}
    \alpha = \nu\Omega_{\rm K}/c_{\rm s}^2,
    \label{eq:alphavis}
\end{equation}
where $\nu$ is the gas viscosity and $c_{\rm s}$ is the sound speed. As a result, the gas surface density is comparable to that of the minimum mass solar nebula. The scale height of the disc $H_{\rm g}$ is a function of the temperature $T$ and the sound speed $c_{\rm s}$ via $H_{\rm g}=c_{\rm s}/\Omega_{\rm K}$, where $c_{\rm s}=(\gamma k_{\rm B}T/\mu m_{\rm p})$. The heat capacity ratio $\gamma$ is set to 7/5, $k_{\rm B}$ is the Boltzmann constant, $\mu=2.3$ is the mean atomic mass of the gas and $m_{\rm p}$ is the mass of the proton. In the simulations, the gas accretion rate varies with time according to \cite{Hartmannetal1998}
\begin{equation}
    \log \left(\frac{\dot{M}_*}{M_{\odot}~{\rm yr}^{-1}}\right) = -8-\frac{7}{5}\log\left(\frac{t+t_{\rm off}}{1~{\rm Myr}}+0.1\right)
\end{equation}
where the extra 0.1~Myr was added to avoid the logarithmic singularity \citep{Bitschetal2015}.\\

The temperature in different regions of the disc is predominantly dictated by the heating source. In general, viscous heating dominates the inner region close to the star while stellar irradiation has a stronger influence in the far away regions. At the midplane, the disc temperature is given by $T = \max\left(T_{\rm vis},T_{\rm irr}\right)$, where $T_{\rm vis}$ and $T_{\rm irr}$ are temperatures in the viscous region and irradiative region, respectively. The empirically fitted expressions for $T_{\rm vis}$ and $T_{\rm irr}$ based on the constant-opacity disc model of \citet{GaraudLin2007} are
\begin{equation}
    \begin{aligned}
    T_{\rm vis} &= T_{0v} \alpha_3^{-1/5} \dot{M}_{*8}^{2/5}
    \left(\frac{r}{1\,{\rm au}}\right)^{-9/10}~{\rm K}, \\
    T_{\rm irr} &= 150 \left(\frac{r}{1\,{\rm au}}\right)^{-3/7}~{\rm K},
    \label{eq:T_visirr}
    \end{aligned}    
\end{equation}
where $T_{0v}$ is the initial temperature at 1~au and $r$ is the distance to the Sun. The power exponents in the equation are derived analytically. We additionally defined the following normalised parameters
\begin{align}
    \alpha_3 &\equiv \frac{\alpha}{10^{-3}}, \\
    \dot{M}_{*8} &\equiv \frac{\dot{M}_*}{10^{-8}~M_\odot~{\rm yr}^{-1}}.
\end{align}
With the disc temperature profiles defined, the reduced disc scale height $h_{\rm g}=H_{\rm g}/r$ can thus be computed as
\begin{equation}
    \begin{aligned}
    h_{\rm g,vis} &= 0.034 \left(\frac{T_{0v}}{200~{\rm K}} \right)^{1/2} \alpha_3^{-1/10}\dot{M}_{*8}^{1/5} 
    \left(\frac{r}{1~{\rm au}}\right)^{1/20}, \\
    h_{\rm g,irr} &= 0.029 \left(\frac{r}{1~{\rm au}}\right)^{2/7}.
    \label{eq:h_visirr}    
    \end{aligned}
\end{equation}
The actual reduced scale height of the disc is $h_{\rm g} = \max(h_{\rm g,vis},h_{\rm g,irr})$. This gives the disc a flaring shape in the irradiative region and a constant scale height in the viscous region. Equations~\ref{eq:dotmstar},~\ref{eq:T_visirr} and~\ref{eq:h_visirr} are then combined to compute the gas surface density in the viscous and irradiative regions
\begin{equation}
    \begin{aligned}
    \Sigma_{\rm g,vis} &= 1320 \left(\frac{T_{0v}}{200~{\rm K}} \right)^{-1} \alpha_3^{-4/5} \dot{M}_{*8}^{3/5}
    \left(\frac{r}{1\,{\rm au}}\right)^{-3/5}{\rm g~cm^{-2}}, \\
    \Sigma_{\rm g,irr} &= 1785 \alpha_3^{-1} \dot{M}_{*8}
    \left(\frac{r}{1\,{\rm au}}\right)^{-15/14}{\rm g~cm^{-2}}.
    \label{eq:Sigma_visirr}    
    \end{aligned}
\end{equation}
The corresponding pressure gradient $\dv*{\ln P}{\ln r}$ in the viscous and irradiative regimes are
\begin{equation}
    \begin{aligned}
    \left( \dv{\ln P}{\ln r} \right)_{\rm vis}  = -\frac{51}{20}, \\
    \left( \dv{\ln P}{\ln r} \right)_{\rm irr}  = -\frac{39}{14}.
    \label{eq:lnplnr_visirr}
    \end{aligned}
\end{equation}
Finally, the boundary between the viscous and irradiative region, determined by $T_{\rm vis} = T_{\rm irr}$, occurs at
\begin{equation}
    \begin{aligned}
    r_{\rm vis/irr} &= \left(\frac{T_{0v}}{150\,{\rm K}} \right)^{70/33} \alpha_3^{-14/33}\dot{M}_{*8}^{28/33} \\ 
    &\approx 1.84~\alpha_3^{-14/33}\dot{M}_{*8}^{28/33}~{\rm au}.
    \label{eq:r_vis_irr}
    \end{aligned}
\end{equation}
%%%%%%%%%%%%%%%%%%%%%%%%%%%%%%%%%%%

%%%%%%%%%%%%%%%%%%%%%%%%%%%%%%%%%%%
\section{Disc-induced orbital evolution}
\label{sec:appendix_migration}
The gas disc exerts torques and tidal forces on all the planetary bodies (with mass $M$) embedded in the disc which result in a combined effect of radial migration and the damping of the orbital eccentricity $e$ and orbital inclination $I$. Low-mass planetary bodies, i.e., those that are unable to clear the surrounding gas, experience type I migration \citep{Tanakaetal2002} while massive bodies experience type II migration \citep{LinPapaloizou1986}. The computation of the torques and the direction of migration is based on the prescriptions of \citet{ColemanNelson2014}. The normalised torque is given by
\begin{equation}
 \frac{\gamma\Gamma}{\Gamma_0} = \frac{\Gamma_{\rm C}}{\Gamma_0}F_{\rm C} + \frac{\Gamma_{\rm L}}{\Gamma_0}F_{\rm L}
 \label{eq:torque}
\end{equation}
where $\Gamma_{\rm C}$ and $\Gamma_{\rm L}$ are the corotation and 
Lindblad torques respectively \citep{Paardekooperetal2011} and 
\begin{equation}
    \Gamma_0= \left( \frac{M}{M_*} \right)^2 h_{\rm g}^{-2} \Sigma_{\rm g}\Omega_{\rm K}^2    
\end{equation}
is a normalisation constant. The factors $F_{\rm L}$ and $F_{\rm C}$ are defined as \citep{CresswellNelson2008,FendykeNelson2014}
\begin{equation}
    \begin{aligned}
    \ln F_{\rm C} &= -\frac{e}{e_{\rm f}}, \\
    \frac{1}{F_{\rm L}} &= P_e + {\rm sign} (P_{\rm e})(0.07\hat{I}+0.085\hat{I}^4-0.08\hat{e}^2\hat{I}^2),
    \label{eq:fcfl}
    \end{aligned}
\end{equation}
where $e_{\rm f}= 0.01+{\textstyle \frac{1}{2}}h_{\rm g}$,~$\hat{e}=e/h_{\rm g}$,~$\hat{I} = \sin(I)/h_{\rm g}$ and 
\begin{equation*}
    P_{\rm e} = \frac{1+\left(0.444\hat{e}\right)^{1/2}+\left(0.352\hat{e}\right)^6}{1-\left(0.495\hat{e}\right)^4}.
\end{equation*}
The eccentricity damping timescale $\tau_e = -e/\dot{e}$ is computed as
\begin{equation}
    \tau_e = 1.282t_{\rm wav}\left(1-0.14\hat{e}^2+0.06\hat{e}^3+0.18\hat{e}^2\hat{I}^2\right),
    \label{eq:taue}
\end{equation}
where the wave timescale is \citep{TanakaWard2004}
\begin{equation}
    t_{\rm wav} = \left(\frac{M_*}{M}\right)\left(\frac{M_*}{\Sigma_{\rm g} r^2}\right)h_{\rm g}^4 \Omega_{\rm K}^{-1}.
\end{equation}
The inclination damping timescale $\tau_{I} = -I/(dI/dt)$ is 
\begin{equation}
    \tau_I = 1.838t_{\rm wav}(1-0.30\hat{I}^2+0.24\hat{I}^3+0.14\hat{e}^2\hat{I}^2).
\end{equation}
For planetary bodies in the type I migration regime, the `migration' timescale is defined as $\tau_{\rm mig,1} = -L/\Gamma$ where $L$ and $\Gamma$ are the angular momentum and the torque. It is computed as
\begin{equation}
\label{eq:tau_mig12}
    \tau_{\rm mig,1} = -\frac{t_{\rm wav}\Gamma_{0}}{h_{\rm g}^2\Gamma}
\end{equation}
in the simulations. For bodies in the type II regime, the migration timescale is computed as
\begin{equation}
    \tau_{\rm mig,2} = \frac{2r^2}{3v}\max\left(1,\frac{M}{2\pi\Sigma_{\rm g}r^2} \right),
\end{equation}
where $r$ is the orbital distance of the planet and $v$ is its orbital speed. The first term in the parentheses corresponds to the case when the planet is less massive than the disc interior to its orbit and the migration occurs on the disc's viscous evolution timescale of $\tau_{\rm mig,2} \simeq (2/3)(r^2/v)$ \citep{LinPapaloizou1986}, whereas the second term corresponds to the opposite case where the planet is more massive and thus the migration occurs on a longer timescale $\tau_{\rm mig,2} \simeq M/\dot{M}_*$ \citep{HasegawaIda2013}.
%%%%%%%%%%%%%%%%%%%%%%%%%%%%%%%%%%%

%%%%%%%%%%%%%%%%%%%%%%%%%%%%%%%%%%%
\section{Gas envelope accretion for massive planets}
\label{sec:appendix_gasaccretion}
Gas accretion commences for planetary bodies that fulfil two criteria: (1) they have reached a critical core mass, and (2) the accretion rate of solids onto said planetary core is sufficiently low that it does not affect the cooling and contraction of the gas envelope. The critical core mass of a planetary body is given by \citep{Ikomaetal2000}
\begin{equation}
    M_{\rm crit} = 10\left( \frac{\dot{M}_{\rm core}}{10^{-6}M_{\rm E}~{\rm yr}^{-1}} \right)^{1/4}M_{\rm E},
\end{equation}
where the core accretion rate is equal to the pebble accretion rate of the planetary body $\dot{M}_{\rm core} = \dot{M}$ (see equation~\ref{eq:dotmpl}). The collapse and accretion of the gas envelope then proceeds on the Kelvin-Helmholtz timescale \citep{Ikomaetal2000} according to
\begin{equation}
    \tau_{\rm KH} = 10^9 \left( \frac{M}{M_{\rm E}} \right)^{-3}~{\rm yr}.
\end{equation}
This process of runaway gas accretion is bounded by the gas accretion rate throughout the disc $\dot{M}_*$. Gas accretion ceases when the Hill radius of the planet is approximately equal to that of the disc scale height. Taking all these effects into account, the gas accretion rate onto the planetary core is written as
\begin{equation}
    \dot{M}_{\rm g} = \min \left[ \frac{M}{\tau_{\rm KH}},\dot{M}_* \exp\left( -\frac{M}{M_{\rm H}} \right) \right],
\end{equation}
where $M_{\rm H} = 3(h_{\rm g}/R)^3M_*$ is the required planetary mass for its Hill radius to be comparable to the disc scale height. The gas accretion rate is limited to the Bondi accretion rate of
\begin{equation}
    \dot{M}_{\rm g,B} = \frac{4\pi\rho_{\rm g}G^2M^2}{c_{\rm s}^3},
\end{equation}
where $\rho_{g} \sim \Sigma_{g}/(\sqrt{2\pi}H_{g})$ is the gas disc's density and $G$ is the gravitational constant. We keep track of the amount of mass in solids and in gas accreted by the planet.
%%%%%%%%%%%%%%%%%%%%%%%%%%%%%%%%%%%

%%%%%%%%%%%%%%%%%%%%%%%%%%%%%%%%%%%
\section{Alternative simulation set up for low-mass discs}
\label{sec:appendix_lowmassdisc}
We carried out some additional simulations for low-mass discs with planetesimal diameter ranging from 100 to 1200~km. We have $N_{\rm plt} \sim 20$ for $M_{\rm disc,i} = 10^{-4}~M_{\rm E}$ and $N_{\rm plt} \sim 150$ for $M_{\rm disc,i} = 10^{-3}~M_{\rm E}$. We simulated with and without fragmentation but turned off migration effects for the sake of simplicity.

\begin{figure}
    \centering
	\includegraphics{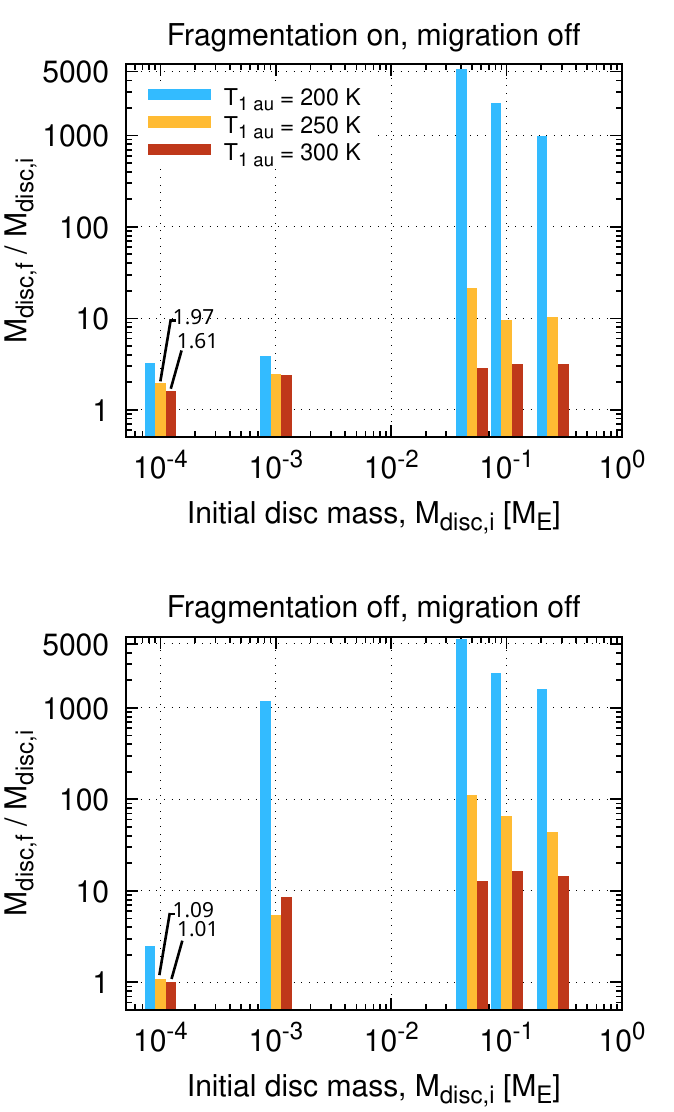}
    \caption{Outcome of additional simulations for $M_{\rm disc,i} = 10^{-4}~{\rm and}~10^{-3}~M_{\rm E}$ with planetesimal diameter range changed to $100~{\rm km} \leq D \leq 1200~{\rm km}$. The overall mass increase in the discs are higher than that shown in Fig.~\ref{fig:pebacc_mfmi_vs_mdisc} because most of the mass in the disc is in a few large planetesimals. The results for discs with $M_{\rm disc,i} \geq 0.05~M_{\rm E}$ are the same as in Fig.~\ref{fig:pebacc_mfmi_vs_mdisc}.}
    \label{fig:pebacc_mfmi_vs_mdisc_extra}
\end{figure}

In Fig.~\ref{fig:pebacc_mfmi_vs_mdisc_extra} we show the outcome of the additional simulations. We find that the mass increase in the low-mass discs is higher than what was reported in Fig.~\ref{fig:pebacc_mfmi_vs_mdisc} and is true for both discs $(M_{\rm disc,i} = 10^{-4}~{\rm and}~10^{-3}~M_{\rm E})$ and all disc temperatures (200~K, 250~K and 300~K). This result is not unexpected as the mass increase would be dominated by one (or a few) large planetesimal(s), which would then go on to scatter the other planetesimals or reach its pebble isolation mass and then stop the flow of pebbles to the other planetesimals.

The disadvantage of this setup is that the outcome is more sensitive to stochastic effects (one large planetesimal growing substantially and dominating the mass of the disc) and thus it is difficult to determine the amount of growth onto small planetesimals and the influence of fragmentation effects; with our setup in the main paper such stochastic effects become more deterministic because the largest planetesimals are generally smaller and the uneven growth is inhibited. Furthermore, with this setup the chances of finding a configuration that satisfies cosmochemical constraints (<10\% accretion) are lower.
%%%%%%%%%%%%%%%%%%%%%%%%%%%%%%%%%%%

%%%%%%%%%%%%%%%%%%%%%%%%%%%%%%%%%%%
\section{Distribution of solids in discs with final mass greater than 2 Earth masses}
\label{sec:appendix_2Earthmass}
\begin{figure*}
    \centering
	\includegraphics[width=0.8\textwidth]{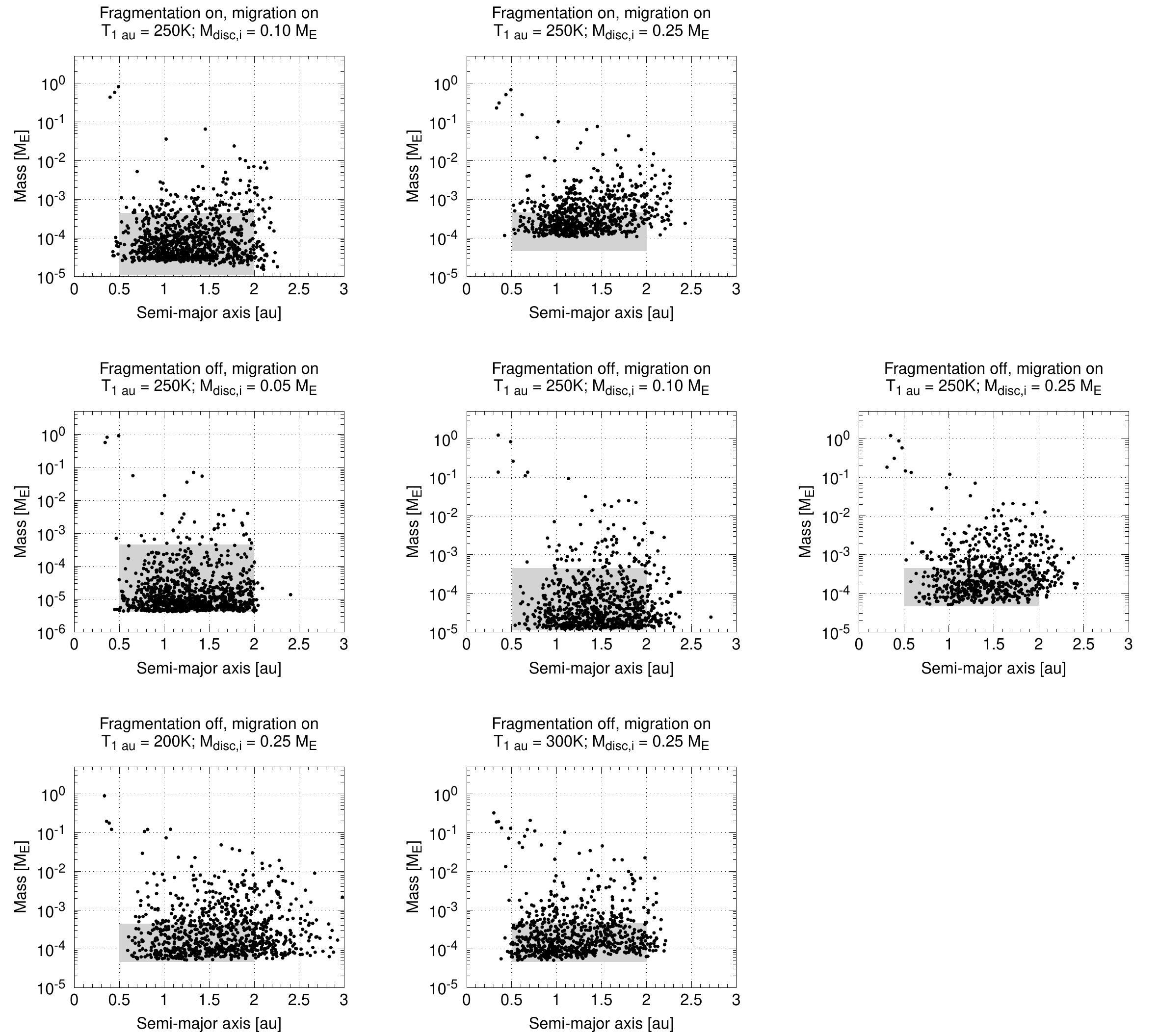}
    \caption{Distribution of mass in the disc at the end of the simulations for cases where the total mass is greater than $2~M_{\rm E}$. The grey shaded region represents the initial mass and semi-major axis distributions of the planetesimals. Most of the mass is in the few objects closer than 0.6~au with Mars' mass or greater.}
    \label{fig:pebacc_2Earthmass}
\end{figure*}

One of the important questions to ask is if the final amount of mass in the solid disc after 4~Myr is sufficient to form the terrestrial planets in the next stage of the planet formation process. We tabulate the total mass in the disc within 3.0~au at the end of the simulations with different initial conditions and migration effects included. The minimum mass of solids in the disc required to form the terrestrial planets is $2~M_{\rm E}$. We find 7 sets of initial conditions that fulfil this requirement (Table~\ref{fig:pebacc_2Earthmass}). We show in Fig.~\ref{fig:pebacc_2Earthmass}  the distribution of solids at 4~Myr for the cases where the final disc mass is greater than $2~M_{\rm E}$. The results appear to suggest that simulations without fragmentation effects are better initial conditions to generate enough mass to form the terrestrial planets. However, as we only ran 1 simulation for each initial condition, the `good' combination of parameters need to be tested more rigorously to determine if they are indeed favourable initial conditions to produce enough mass in the disc for the subsequent stage of planet formation. For the discs with $M_{\rm disc,i} \leq 1\times10^{-3}~M_{\rm E}$ we do not expect them to be able to form the terrestrial planets because the planetesimals would not grow by much. They serve to test the robustness of the analytical result in Fig.~\ref{fig:pebacc_contour} that planetesimals smaller than a critical size are very inefficient at accreting pebbles.

\begin{table}
	\centering
	\caption{Initial conditions of simulations where the final mass in the solid disc is greater than $2~M_{\rm E}$}
    \label{tab:pebacc_2Earthmass}
    \begin{tabular}{CBBB}
    \toprule
    Fragmentation & Initial disc temperature $T_{\rm 1~au}~({\rm K})$ & Initial solid disc mass $M_{\rm disc,i}~(M_{\rm E})$  & Final solid disc mass $M_{\rm disc,f}~(M_{\rm E})$\\ 
    \midrule
    On  & 250 & 0.10 & 2.21 \\
    On  & 250 & 0.25 & 2.80 \\
    Off & 200 & 0.25 & 2.72 \\
    Off & 250 & 0.05 & 2.63 \\
    Off & 250 & 0.10 & 3.11 \\
    Off & 250 & 0.25 & 4.25 \\
    Off & 300 & 0.25 & 2.44 \\    
    \bottomrule
    \end{tabular}
\end{table}
%%%%%%%%%%%%%%%%%%%%%%%%%%%%%%%%%%%%%%%%%%%%%%%%%%

% Don't change these lines
\bsp	% typesetting comment
\label{lastpage}
\end{document}